\documentclass[12pt]{article}
\begin{document}
\pagestyle{empty}

\begin{center} 
{\Large {\bf Sleptons  Masses in the 
Minimal Supersymmetric $SU(3)_{C}\times SU(3)_{L}\times U(1)_{N}$ Model.}}
\end{center}

\begin{center}
M. C. Rodriguez  \\
{\it Grupo de F{\'{\i}}sica Te\'{o}rica e Matem\'{a}tica F\'{\i}sica \\
Departamento de F\'{\i}sica  \\
Universidade Federal Rural do Rio de Janeiro - UFRRJ \\
BR 465 Km 7, 23890-000, Serop\'{e}dica - RJ \\
Brasil}
\end{center}

\date{\today} 

\begin{abstract}
We present the mass spectrum of the sleptons and squarks in the  Minimal Supersymmetric 
$SU(3)_{C}\times SU(3)_{L}\times U(1)_{N}$  Model, when all the usual neutral scalars fields obtain vacuum 
expectation values. In this model, we can define $R$-Parity conservation, where our superpotential respect 
the invariance of the quantum number ${\cal F} \equiv B+L$, where $B$ is the baryon number while $L$ is 
the total lepton number. 
\end{abstract}

PACS number(s): 12.60. Cn, 12.60. Jv


\section{Introduction}

Models with the gauge symmetry \cite{singer,ppf,331rh,Pleitez:1992xh,Pleitez:1993gc} 
\begin{equation}
SU(3)_{C} \times SU(3)_{L} \times U(1)_{N}
\label{331symmetry}
\end{equation} 
are known as $331$ for short and offer various possibilities and they are 
interesting possibilities for the physics at the TeV scale 
\cite{Pleitez:1994pu,Ponce:2001jn,Diaz:2004fs,Pleitez:2021abk,Long:2024gyy}. It is 
a subgroup of unification group $E_{6}$ \cite{Sanchez:2001ua} and it is also 
an $SU(6) \times U(1)_{X}$  \cite{Martinez:2001mu}, $SU(8)$ \cite{Martinez:2011iq} 
and $SU(15)$ \cite{Frampton:1989fu} subgroups. This fact allow us to explain the 
hierarchy problem because in this case exist a new scale and the chain of breaking 
is written as GUT $\rightarrow$ 331 $\rightarrow$ 321.

In this model the charge operator is
\begin{eqnarray}
\frac{Q^{T}}{e}&=& T^{3}- \vartheta T^{8}+N \,\ I_{3 \times 3}, \,\ T^{a}= \frac{\lambda^{a}}{2}, 
\label{co}
\end{eqnarray}
where $\lambda^{a}$ are the Gell-Mann matrices. We can introduce the fermions in the following way \cite{Diaz:2004fs}
\begin{itemize}
\item[1-)] Triplet $(3)$
\begin{itemize}
\item[a-)] Leptons
\begin{eqnarray}
L_{L}&\equiv& 
\left( \begin{array}{c} 
\nu_{L} \\ 
l_{L} \\
E^{Q^{T}_{l}}          
\end{array} \right) \sim ({\bf1},{\bf3}, N^{L}_{l}), \,\
N^{L}_{l}=- \frac{1}{2}- \frac{\vartheta}{2 \sqrt{3}}, \,\
Q^{T}_{l}=- \frac{1}{2}- \frac{\sqrt{3} \vartheta}{2}, \nonumber \\
\label{leptonstripletogeral}
\end{eqnarray}
\item[b-)] Quarks
\begin{eqnarray}
Q_{L}&\equiv& 
\left( \begin{array}{c} 
u_{L} \\ 
d_{L} \\
J^{Q^{T}_{J}}          
\end{array} \right) \sim ({\bf3},{\bf3}, N^{L}_{q}), \,\
N^{L}_{q}= \frac{1}{6}- \frac{\vartheta}{2 \sqrt{3}}, \,\
Q^{T}_{J}= \frac{1}{6}- \frac{\sqrt{3} \vartheta}{2}, \nonumber \\ 
\label{quarkstripletogeral}
\end{eqnarray}
\item[c-)]Covariant derivative
\begin{equation}
{\cal D}^{T}_{m} \psi_{L}= \partial_{m}\psi_{L}- 
\imath g \left( V^{a}_{m}T^{a} \right) \psi_{L}- 
\imath g^{\prime} N_{\psi_{L}}V_{m}\psi_{L}.
\label{derivadacovariantetripleto}
\end{equation}
\end{itemize}
\end{itemize}

They are interesting possibilities for the physics at the TeV scale due the 
following nice aspects 
\begin{enumerate}
\item The family number must be a multiple of three in order to cancel 
anomalies \cite{ppf,331rh};
\item Why $\sin^{2} \theta_{W}<\frac{1}{4}$, see Eq.(\ref{tinthetaw2}), is 
observed \cite{ppf}.
\item It is the simplest model that includes bileptons of both types: Scalars and 
Vectors ones \cite{Cuypers:1996ia}, see our Eq.(\ref{efe});
\item The models have a scalar sector similar to the Two Higgs Doublets Models 
\cite{Liu:1993gy,Liu:1993fwa,Montero:1999mc,DeConto:2015eia};
\item The SM plus two scalar doublets $\Phi_{\eta}, \Phi_{\rho}$, see Eq.(\ref{22HM}), and 
$\Phi_{S}$ \cite{Machado:2010uc};
\item A singly charged singlet scalar $\eta^{+}_{2}$, inducing Majorana masses for neutrino at 1-loop level \cite{Zee:1980ai,Machado:2017flo};
\item A doubly charged singlet $H^{--}_{2}$ has been considered  for generating Majorana neutrino masses at the 2-loop level \cite{Zee:1985id,Babu:1988ki};
\item A complex triplet $T$, see Eq.(\ref{22HM}), generating the type-II see 
saw model \cite{Gelmini:1980re,Schechter:1980gr,Cheng:1980qt};
\item The scalar doublet $\Phi_{\chi}$, which is part of a triplet of $SU(3)$, has also been considered as extension of the SM~\cite{Aoki:2011yk}, for instance, in a new scotogenic model~\cite{Puerta:2021hfl}. 
\item The models predict the quantization of the electric charge 
\cite{deSousaPires:1998jc,Dong:2005ebq};
\item It solves the strong $CP$-Problem \cite{pal1};
\item The model has several sources of $CP$-Violation 
\cite{Montero:1998nf,Montero:1998yw,Montero:2005yb}.
\end{enumerate}

\section{Minimal Supersymmetric $331$ Model.}
\label{msusy1}

We introduce the following chiral superfield for the leptons in the triplet representation of 
$SU(3)_{L}$ and we can decompose those fields in 
\begin{equation}
\left( SU(3)_{C} \times SU(3)_{L}\times U(1)_{N} \right) \supset 
\left( SU(3)_{C} \times SU(2)_{L}\times U(1)_{Y} \right),
\label{simetria331}
\end{equation} 
in the following way 
\cite{331susy1,mcr,Rodriguez:2022hsj,Rodriguez:2024rvf,Rodriguez:2025vea}
\begin{eqnarray}
\hat{L}_{iL}&=&\left( 
\begin{array}{c}
\hat{l}^{D}_{iL} \\ 
\hat{l}^{c}_{iL}         
\end{array} 
\right) \sim \left( {\bf 1},{\bf 3},0 \right),  \,\ i=1,2,3, 
\nonumber \\
\hat{l}^{D}_{iL}&=&\left( 
\begin{array}{c}
\hat{\nu}_{iL} \\ 
\hat{l}_{iL}         
\end{array} 
\right) \sim \left( {\bf 1},{\bf 2},-1 \right),  \,\
\hat{l}^{c}_{iL}\sim \left( {\bf 1},{\bf 1},2 \right).
\label{quiralLMSUSY331}
\end{eqnarray}
The particle content of each chiral superfield and anti-chiral supermultiplet is 
presented in the Tabs.(\ref{lfermionnmssm},\ref{rfermionnmssm}), respectively. 

\begin{table}[h]
\begin{center}
\begin{tabular}{|c|c|c|}
\hline 
$\mbox{ Chiral Superfield} $ & $\mbox{ Fermion} $ & $\mbox{ Scalar} $ \\
\hline
$\hat{L}_{iL}=( \hat{l}^{D}_{iL}, \hat{l}^{c}_{iL})^{T}$ & 
$L_{iL}=( l^{D}_{iL},l^{c}_{iL})^{T}$ & 
$\tilde{L}_{iL}=( \tilde{l}^{D}_{iL}, \tilde{l}^{c}_{iL})^{T}$ \\
\hline
$\hat{Q}_{3L}= \left( \hat{q}^{D}_{3L}, \hat{J}_{L} \right)^{T}$ & 
$Q_{3L}=( q^{D}_{3L},J_{L})^{T}$ & 
$\tilde{Q}_{3L}=( \tilde{q}^{D}_{3L}, \tilde{J}_{L})^{T}$ \\ 
\hline
$\hat{\eta} = \left( \hat{\Phi}_{\eta}, \hat{\eta}^{+}_{2} \right)^{T}$ &
$\tilde{\eta} = \left( \tilde{\Phi}_{\eta}, \tilde{\eta}^{+}_{2} \right)^{T}$ &
$\eta = \left( \Phi_{\eta}, \eta^{+}_{2} \right)^{T}$ \\
\hline
$\hat{\rho} = \left( \hat{\Phi}_{\rho}, \hat{\rho}^{++} \right)$ &
$\tilde{\rho} = \left( \tilde{\Phi}_{\rho}, \tilde{\rho}^{++} \right)$ &
$\rho = \left( \Phi_{\rho}, \rho^{++} \right)$ \\
\hline
$\hat{\chi} = \left( \hat{\Phi}_{\chi}, \hat{\chi}^{0} \right)$ &
$\tilde{\chi} = \left( \tilde{\Phi}_{\chi}, \tilde{\chi}^{0} \right)$ &
$\chi = \left( \Phi_{\chi}, \chi^{0} \right)$ \\
\hline
$\hat{S}^{\prime}= \left( 
\begin{array}{cc} 
\hat{T}^{\prime} & \frac{\hat{\Phi}^{\prime}_{S}}{\sqrt{2}} \\ 
\frac{\hat{\Phi}^{\prime T}_{S}}{\sqrt{2}} & \hat{H}^{\prime ++}_{2}          
\end{array} \right)$ &
$\tilde{S}^{\prime}= \left( 
\begin{array}{cc} 
\tilde{T}^{\prime} & \frac{\tilde{\Phi}^{\prime}_{S}}{\sqrt{2}} \\ 
\frac{\tilde{\Phi}^{\prime T}_{S}}{\sqrt{2}} & \tilde{H}^{\prime ++}_{2}          
\end{array} \right)$ & 
$S^{\prime}= \left( 
\begin{array}{cc} 
T^{\prime} & \frac{\Phi^{\prime}_{S}}{\sqrt{2}} \\ 
\frac{\Phi^{\prime T}_{S}}{\sqrt{2}} & H^{\prime ++}_{2}          
\end{array} \right)$ \\
\hline
\end{tabular}
\end{center}
\caption{Particle content in the chiral superfields in MSUSY331 
and we neglected the color indices and $i=1,2,3$.}
\label{lfermionnmssm}
\end{table} 

\begin{table}[h]
\begin{center}
\begin{tabular}{|c|c|c|}
\hline 
$\mbox{ Anti-Chiral Superfield} $ & $\mbox{ Fermion} $ & $\mbox{ Scalar} $  \\
\hline
$\hat{l}^{c}_{iL}$ & $l^{c}_{iL}\equiv \bar{l}_{iR}$ & $\tilde{l}^{c}_{iL}$ \\
\hline
$\hat{Q}_{\alpha L}=( \hat{q}^{D}_{\alpha L}, \hat{j}_{\alpha L})^{T}$ & 
$Q_{\alpha L}=( q^{D}_{\alpha L},j_{\alpha L})^{T}$ & 
$\tilde{Q}_{\alpha L}=
( \tilde{q}^{D}_{\alpha L}, \tilde{j}_{\alpha L})^{T}$ \\
\hline
$\hat{u}^{c}_{iL}$ & $u^{c}_{iL}\equiv \bar{u}_{iR}$ & 
$\tilde{u}^{c}_{iL}$ \\ 
\hline
$\hat{d}^{c}_{iL}$ & 
$d^{c}_{iL}\equiv \bar{d}_{iR}$ & 
$\tilde{d}^{c}_{iL}$ \\ 
\hline
$\hat{j}^{c}_{\alpha L}$ & 
$j^{c}_{\alpha L}\equiv \bar{j}_{\alpha R}$ & 
$\tilde{j}^{c}_{\alpha L}$   \\
\hline
$\hat{J}^{c}_{L}$ & 
$J^{c}_{L}\equiv \bar{J}_{R}$ & 
$\tilde{J}^{c}_{L}$   \\
\hline
$\hat{\eta}^{\prime} = \left( \hat{\Phi}^{\prime}_{\eta}, \hat{\eta}^{\prime -}_{2} \right)^{T}$ &
$\tilde{\eta}^{\prime} = \left( \tilde{\Phi}^{\prime}_{\eta}, \tilde{\eta}^{\prime -}_{2} \right)^{T}$ &
$\eta^{\prime} = \left( \Phi^{\prime}_{\eta}, \eta^{\prime -}_{2} \right)^{T}$ \\
\hline
$\hat{\rho}^{\prime}= \left( \hat{\Phi}^{\prime}_{\rho}, \hat{\rho}^{\prime --} \right)^{T}$ &
$\tilde{\rho}^{\prime}= \left( \tilde{\Phi}^{\prime}_{\rho}, \tilde{\rho}^{\prime --} \right)^{T}$ & 
$\rho^{\prime} = \left( \Phi^{\prime}_{\rho}, \rho^{\prime --} \right)^{T}$ \\
\hline
$\hat{\chi}^{\prime}= \left( \hat{\Phi}^{\prime}_{\chi}, \hat{\chi}^{\prime 0} \right)^{T}$ &
$\tilde{\chi}^{\prime}= \left( \tilde{\Phi}^{\prime}_{\chi}, \tilde{\chi}^{\prime 0} \right)^{T}$ &
$\chi^{\prime} = \left( \Phi^{\prime}_{\chi}, \chi^{\prime 0} \right)^{T}$ \\
\hline
$\hat{S}= \left( 
\begin{array}{cc} 
\hat{T} & \frac{\hat{\Phi}_{S}}{\sqrt{2}} \\ 
\frac{\hat{\Phi}^{T}_{S}}{\sqrt{2}} & \hat{H}^{--}_{2}          
\end{array} \right)$ & 
$\tilde{S}= \left( 
\begin{array}{cc} 
\tilde{T} & \frac{\tilde{\Phi}_{S}}{\sqrt{2}} \\ 
\frac{\tilde{\Phi}^{T}_{S}}{\sqrt{2}} & \tilde{H}^{--}_{2}          
\end{array} \right)$ &
$S= \left( 
\begin{array}{cc} 
T & \frac{\Phi_{S}}{\sqrt{2}} \\ 
\frac{\Phi^{T}_{S}}{\sqrt{2}} & H^{--}_{2}          
\end{array} \right)$ \\
\hline
\end{tabular}
\end{center}
\caption{Particle content in the anti-chiral superfields in MSUSY331 and 
$\alpha , \beta =1,2$, $i=1,2,3$.}
\label{rfermionnmssm}
\end{table} 

The quarks are
\begin{eqnarray}
\hat{Q}_{\alpha L}&=&\left( 
\begin{array}{c}
\hat{q}^{D}_{\alpha L} \\ 
\hat{j}_{\alpha L}         
\end{array} 
\right) \sim \left( {\bf 3},{\bf 3},- \frac{1}{3} \right),  \,\ \alpha =1,2, \nonumber \\  
\hat{q}^{D}_{\alpha L}&=&\left( 
\begin{array}{c}
\hat{d}_{\alpha L} \\ 
\hat{u}_{\alpha L}         
\end{array} 
\right) \sim \left( {\bf 3},{\bf 2}, \frac{1}{3} \right),  \,\
\hat{j}_{\alpha L}\sim \left( {\bf 3},{\bf 1},- \frac{8}{3} \right), \nonumber \\
\hat{Q}_{3L}&=&\left( 
\begin{array}{c}
\hat{q}^{D}_{3L} \\ 
\hat{J}_{L}         
\end{array} 
\right) \sim \left( {\bf 3},{\bf \bar{3}}, \frac{2}{3} \right), 
\nonumber \\
\hat{q}^{D}_{3L}&=&\left( 
\begin{array}{c}
\hat{u}_{3L} \\ 
\hat{d}_{3L}         
\end{array} 
\right) \sim \left( {\bf 3},{\bf 2}, \frac{1}{3} \right),  \,\
\hat{J}_{L}\sim \left( {\bf 3},{\bf 1}, \frac{10}{3} \right).
\label{quiral12QMSUSY331in321}
\end{eqnarray}

We have also to introduce the following singlets
\begin{eqnarray}
\hat{u}^{c}_{\alpha L} &\sim& 
\left( {\bf \bar{3}},{\bf 1},- \frac{2}{3} \right),  \,\
\hat{d}^{c}_{\alpha L} \sim 
\left( {\bf \bar{3}},{\bf 1}, \frac{1}{3} \right),  \,\
\hat{j}^{c}_{\alpha L} \sim 
\left( {\bf \bar{3}},{\bf 1}, \frac{4}{3} \right), \nonumber \\
\hat{u}^{c}_{3L} &\sim& 
\left( {\bf \bar{3}},{\bf 1},- \frac{2}{3} \right),  \,\
\hat{d}^{c}_{3L} \sim 
\left( {\bf \bar{3}},{\bf 1}, \frac{1}{3} \right),  \,\
\hat{J}^{c}_{L} \sim 
\left( {\bf \bar{3}},{\bf 1},- \frac{5}{3} \right).
\label{quiral12QMSUSY331rh}
\end{eqnarray}
The heavy quarks $J,j_{1},j_{2}$ which would imply the existence of hadrons with 
exotic quarks~\cite{DeConto:2017aob}.

The scalars of this model are \cite{331susy1,mcr,Rodriguez:2022hsj,Rodriguez:2024rvf}
\begin{eqnarray}
\eta &=& \left( 
\begin{array}{c} 
\Phi_{\eta} \\ 
\eta^{+}_{2}          
\end{array} \right) 
\sim ({\bf1},{\bf3},0), \,\ 
\rho = \left( 
\begin{array}{c} 
\Phi_{\rho} \\ 
\rho^{++}          
\end{array} \right) 
\sim ({\bf 1},{\bf 3},+1),  \nonumber \\
\Phi_{\eta} &=& \left( 
\begin{array}{c} 
\eta^{0} \\ 
\eta^{-}_{1}          
\end{array} \right) 
\sim ({\bf1},{\bf2},-1),\quad 
\eta^{+}_{2} 
\sim ({\bf 1},{\bf 1},+2),  \nonumber \\
\Phi_{\rho} &=& \left( 
\begin{array}{c} 
\rho^{+} \\ 
\rho^{0}          
\end{array} \right) 
\sim ({\bf 1},{\bf 2},+1), \quad
\rho^{++} \sim ({\bf 1},{\bf 1},+4), \nonumber \\
\chi &=& \left( 
\begin{array}{c} 
\Phi_{\chi} \\ 
\chi^{0}          
\end{array} \right) 
\sim ({\bf 1},{\bf 3},-1), \,\ 
\Phi_{\chi} = \left( 
\begin{array}{c} 
\chi^{-} \\ 
\chi^{--}          
\end{array} \right) 
\sim ({\bf 1},{\bf 2},-3),  \nonumber \\
S&=& \left( 
\begin{array}{cc} 
T & \frac{\Phi_{S}}{\sqrt{2}} \\ 
\frac{\Phi^{T}_{S}}{\sqrt{2}} & H^{--}_{2}          
\end{array} \right)
\sim ({\bf 1},{\bf {\bf \bar{6}}},0), \quad
\chi^{0}\sim ({\bf 1},{\bf 1},0), 
\nonumber \\
\Phi_{S}&=& \left( 
\begin{array}{c} 
h^{+}_{2} \\ 
\sigma^{0}_{2}          
\end{array} \right)
\sim ({\bf 1},{\bf 2},+1), \quad 
H^{--}_{2} \sim ({\bf 1},{\bf 1},+4).
\label{22HM}
\end{eqnarray}
The doublets $\Phi_{\eta}, \Phi_{\rho}$ and $\Phi_{S}$ are similar ones considered 
in \cite{Machado:2010uc}. The singly charged singlet scalar $\eta^{+}_{2}$ induce 
Majorana masses for neutrinos at 1-loop level \cite{Zee:1980ai,Machado:2017flo}. 
The double charged scalar $H^{--}_{2}$ has been considered  for generating 
Majorana neutrino masses at the 2-loop level \cite{Zee:1985id,Babu:1988ki}. The 
triplet $T$ generate the type-II see saw model 
\cite{Gelmini:1980re,Schechter:1980gr,Cheng:1980qt}
\begin{eqnarray}
T&=&\left( \begin{array}{cc} 
\sigma^{0}_{1} & \frac{h^{+}_{1}}{\sqrt{2}} \\ 
\frac{h^{+}_{1}}{\sqrt{2}} & H^{--}_{1} \\
\end{array} \right)
\sim ({\bf 1},{\bf 3},+2).
\label{tripin6^{*}}
\end{eqnarray}

In order to implement supersymmetry, and also at same time to cancel chiral 
anomalies, we must to introduce the following scalars fields \cite{331susy1,mcr}
\begin{eqnarray}
\eta^{\prime} &=& 
\left( 
\begin{array}{c} 
\Phi^{\prime}_{\eta} \\
\eta^{\prime -}_{2}          
\end{array} \right) 
\sim ({\bf1},{\bf \bar{3}},0), \,\
\Phi^{\prime}_{\eta} = \left( 
\begin{array}{c} 
\eta^{\prime 0} \\ 
\eta^{\prime +}_{1}          
\end{array} \right) \sim ({\bf 1},{\bf \bar{2}},+1), \nonumber \\
\rho^{\prime} &=& 
\left( \begin{array}{c} 
\Phi^{\prime}_{\rho} \\
\rho^{\prime --}          
\end{array} 
\right) 
\sim ({\bf1},{\bf \bar{3}},-1), \,\
\Phi^{\prime}_{\rho} = \left( 
\begin{array}{c} 
\rho^{\prime -} \\ 
\rho^{\prime 0}          
\end{array} \right) \sim ({\bf 1},{\bf \bar{2}},+1), \nonumber \\ 
\chi^{\prime} &=& 
\left( \begin{array}{c} 
\Phi^{\prime}_{\chi} \\
\chi^{\prime 0}          
\end{array} \right) 
\sim ({\bf1},{\bf \bar{3}},+1), \,\
\Phi^{\prime}_{\chi} = \left( 
\begin{array}{c} 
\chi^{\prime +} \\ 
\chi^{\prime ++}          
\end{array} \right) \sim ({\bf 1},{\bf \bar{2}},+1), \nonumber \\
S^{\prime}&=& \left( 
\begin{array}{cc} 
T^{\prime} & \frac{\Phi^{\prime}_{S}}{\sqrt{2}} \\ 
\frac{\Phi^{\prime T}_{S}}{\sqrt{2}} & H^{\prime ++}_{2}          
\end{array} \right)
\sim ({\bf 1},{\bf {\bf 6}},0), \,\ 
\Phi^{\prime}_{S}= \left( 
\begin{array}{c} 
h^{\prime -}_{2} \\ 
\sigma^{\prime 0}_{2}          
\end{array} \right) \sim ({\bf 1},{\bf \bar{2}},+1), \nonumber \\ 
T^{\prime}&=&\left( \begin{array}{cc} 
\sigma^{\prime 0}_{1} & \frac{h^{\prime -}_{1}}{\sqrt{2}} \\ 
\frac{h^{\prime }_{1}}{\sqrt{2}} & H^{\prime ++}_{1} \\
\end{array} \right)
\sim ({\bf 1},{\bf \bar{3}},+2). \nonumber \\
\label{3t} 
\end{eqnarray}

In the MSUSY331 we need to introduce the following three vector superfields 
$\hat{V}^{a}_{C}\sim({\bf 8},{\bf 1}, 0)$
\footnote{The gluinos are the superpartner of gluons, and therefore they are in the adjoint representation of $SU(3)$, which is real.}, 
where $a=1,2, \ldots ,8$, $\hat{V}^{a}\sim({\bf 1},{\bf 8}, 0)$, and 
$\hat{V}\sim({\bf 1},{\bf 1}, 0)$. The particle content 
in each vector superfield is presented in the Tab.(\ref{gaugemssm}). 
\begin{table}[h]
\begin{center}
\begin{tabular}{|c|c|c|c|}
\hline 
${\rm{Vector \,\ Superfield}}$ & ${\rm{Gauge \,\ Bosons}}$ & ${\rm{Gaugino}}$ & ${\rm Gauge \,\ constant}$ \\
\hline 
$\hat{V}^{a}_{C}\sim({\bf 8},{\bf 1}, 0)$ & $g^{a}_{m}$ & $\lambda _{C}^{a}$ & $g_{s}$ \\
\hline 
$\hat{V}^{a}\sim({\bf 1},{\bf 8}, 0)$ & $V^{a}_{m}$ & $\lambda^{a}_{A}$ & $g$ \\
\hline
$\hat{V}\sim({\bf 1},{\bf 1}, 0)$ & $V_{m}$ & $\lambda$ & $g^{\prime}$ \\
\hline
\end{tabular}
\end{center}
\caption{\small Particle content in the vector superfields in 
MSUSY331, where $a=1,2, \ldots ,8$.}
\label{gaugemssm}
\end{table}

We denote the $SU(3)_{L}$ gauge bosons by $V^{a}_{m}$ ($a=1,2, \ldots ,8$) and since 
\begin{eqnarray}
\left( {\bf 1}, {\bf 8}, 0 \right) \rightarrow 
\left( {\bf 1},{\bf 3}, 0 \right) \oplus
\left( {\bf 1},{\bf 1}, 0 \right) \oplus  
\left( {\bf 1}, {\bf 2}, 3 \right) \oplus 
\left( {\bf 1}, {\bf 2}, -3 \right),
\end{eqnarray} 
we get a triplet, $V^{1}_{m},V^{2}_{m},V^{3}_{m}$, and also a 
singlet $V^{8}_{m}$ both with $Y=0$, plus the doublet of bileptons, 
$\left( V^{4}_{m},V^{5}_{m} \right)$, its hypercharge is 
$Y= 3$ and $\left( V^{6}_{m},V^{7}_{m} \right)$ with $Y=-3$. The charged gauge bosons of 
this model are \cite{331susy1,mcr}
\begin{eqnarray}
W^{ \pm}_{m}(x)&=&-\frac{1}{\sqrt{2}}(V^{1}_{m}(x) 
\mp \imath V^{2}_{m}(x)), \,\
V^{ \pm}_{m}(x)=-\frac{1}{\sqrt{2}}(V^{4}_{m}(x) 
\pm \imath V^{5}_{m}(x)) \nonumber \\
U^{\pm \pm}_{m}(x) &=&- \frac{1}{\sqrt{2}}(V^{6}_{m}(x) 
\pm \imath V^{7}_{m}(x)).
\label{defcarbosons}
\end{eqnarray}

The neutral gauge bosons, with $Y=0$, are the photon
\begin{eqnarray}
A_{m}&=& \sin \theta_{W} \left( V^{3}_{m}- \sqrt{3}V^{8}_{m} \right) + 
\sqrt{1-4 \sin^{2} \theta_{W}}V_{m},
\label{photondef}
\end{eqnarray}
where $\theta_{W}$ is the Weinberg angle and we have also two massives neutral bosons defined as
\begin{eqnarray}
Z_{m}&=& \cos \theta_{W}V^{3}_{m}+ \sqrt{3}\tan \theta_{W}\sin \theta_{W}V^{8}_{m}- 
\tan \theta_{W}\sqrt{1-4 \sin^{2} \theta_{W}}V_{m}, \nonumber \\
Z^{\prime}_{m}&=& \frac{1}{\cos \theta_{W}}\left[
\sqrt{1-4 \sin^{2} \theta_{W}}V^{8}_{m}+ \sqrt{3}\sin \theta_{W}V_{m} \right].
\label{Z-ZPdef}
\end{eqnarray}
In the MSUSY331, we can write \cite{Rodriguez:2022hsj,Rodriguez:2024rvf}
\begin{equation}
\frac{M^{2}_{Z}}{M^{2}_{W}}= \frac{1+4t^{2}}{1+3t^{2}} \equiv \frac{1}{\cos^{2} \theta_{W}}, \,\ 
t \equiv \frac{g^{\prime}}{g},
\label{tinthetawm331}
\end{equation}
we obtain the famous relationship
\begin{equation}
t^{2}= \frac{\sin^{2} \theta_{W}}{1-4 \sin^{2} \theta_{W}},
\label{tinthetaw2}
\end{equation}
To avoid the loss of the model being perturbative, it allows us to write the following inequality \cite{Ng:1992st}
\begin{eqnarray}
\sin^{2}\theta_{W} \left( M_{Z^{\prime}} \right)< \frac{1}{4}.
\label{muesc}
\end{eqnarray} 

The interaction between the charged bosons with the leptons are given by \cite{mcr,Rodriguez:2010tn}
\begin{eqnarray}
{\cal L}_l^{CC}&=&-\frac{g}{\sqrt{2}}\sum_{l}\left(
\bar{\nu}_{lL}\gamma^{m}V_{\rm PMNS}l_{L}W^{+}_{m}+ 
\bar{l}^{c}_{L}\gamma^{m}U_{V}\nu_{lL} V^{+}_{m}+
\bar{l}^{c}_{L}\gamma^{m} U_{U}l_{L}U^{++}_{m}+hc \right). \nonumber \\
\label{lf}
\end{eqnarray}
The gauge bosons $V^{\pm}$ and $U^{\pm \pm}$ are known as bileptons 
\cite{Cuypers:1996ia}. The $V_{\rm PMNS}$ is the Pontecorvo-Maki-Nakagawa-Sakata 
mixing matrix. There are new mixing matrices given by $U_{V}$ and $U_{U}$. The same 
interactions with the quarks and gauge bosons are
\begin{eqnarray}
{\cal L}_q^{CC}& = & -\frac{g}{2\sqrt{2}}\left[\overline{U}\gamma^m(1 - 
\gamma_5)V_{\rm CKM}DW^+_m  + \overline{U}\gamma^m(1 - \gamma_5)\zeta{\cal 
JV}_m \right. \nonumber \\ &+& \left. 
\overline{D}\gamma^m(1 - \gamma_5)\xi{\cal JU}_m\right] + hc, 
\nonumber \\
\label{lq}
\end{eqnarray}
where we have defined the mass eigenstates in the following way
\begin{eqnarray}
U &=& \left(\begin{array}{c} u \\ c \\ t
\end{array}\right), \quad
D =  \left(\begin{array}{c}
         d \\ s \\ b
\end{array}\right), \quad
{\cal V}_m = \left(\begin{array}{c}
         V^+_m \\ U^{--}_m \\ U^{--}_m\end{array}\right), \nonumber \\
{\cal U}_m & = & \left(\begin{array}{c}
         U^{--}_m \\ V^+_m \\ V^+_m
\end{array}\right), \,\ 
{\cal J}=  \left(\begin{array}{c}
        J_{1} \equiv j_{1} \\ J_{2} \equiv j_{2} \\ J_{3} \equiv J
\end{array}\right).  
\label{maest}
\end{eqnarray}
The $V_{\rm CKM}$ is the usual Cabibbo-Kobayashi-Maskawa (CKM) mixing matrix,  and 
$\xi$ and $\zeta$ are mixing matrices containing new unknown mixing parameters due 
to the presence of the exotic quarks.

The extra neutral vector boson $Z^{\prime}$ interaction with the usual fermions are 
given by the following lagrangian \cite{dumm2}
\begin{eqnarray}
{\cal L}_{Z^{\prime}}&=&- \frac{g}{\cos \theta_{W}} \left[
\sum_{F=L,Q}\bar{F}_{L}g_{L}\gamma^{m}F_{L}+
\sum_{q=u_{i},d_{i},J_{i}}\bar{q}_{R}g_{R}\gamma^{m}q_{R} \right] Z^{\prime}_{m},
\end{eqnarray}
where
\begin{eqnarray}
g_{L}&=& \frac{\sqrt{3}\sin^{2}\theta_{W}}{\sqrt{1-4s^{2}_{W}}}N+ \sqrt{1-4s^{2}_{W}}T_{8}, \nonumber \\
g_{R}&=&\frac{\sqrt{3}\sin^{2}\theta_{W}}{\sqrt{1-4s^{2}_{W}}}N
\end{eqnarray}
The couplings to the leptons, $N=0$\footnote{Review our 
Eqs.(\ref{derivadacovariantetripleto},\ref{simetria331},\ref{quiralLMSUSY331}).}, are leptophobic 
because of the suppression factor 
\begin{equation}
h(t)=\sqrt{1-4s^{2}_{W}},
\end{equation} 
but with some quarks there are enhancements because of the factor $(h(t))^{-1}$ \cite{dumm2}. Therefore, the 
Flavour Changing Neutral Current (FCNC) arise only in the quark sector\footnote{Review 
Tabs.(\ref{lfermionnmssm},\ref{rfermionnmssm})} and detailed analyses have been presented in 
\cite{Buras:2012dp,Buras:2014yna,Buras:2021rdg}.

Those new  vector bosons can induce the following new decays
\begin{eqnarray}
V&\to& u+J, \,\ V \to d+j, \nonumber \\
U&\to& d+J, \,\ U \to u+j, \nonumber \\
Z^{\prime}&\to& \bar{\nu}+ \nu .
\end{eqnarray}
One day, we will need to worry to consider \cite{Pleitez:2021abk} 
\begin{itemize}
\item The search of exotic  quarks or leptons;
\item Exotic baryons;
\item Confusion among neutrinos and antineutrinos.
\end{itemize}

With those fields, we can built the following lagrangian \cite{331susy1}
\begin{eqnarray} 
{\cal L}_{331S} &=& {\cal L}_{SUSY} + {\cal L}_{\mbox{soft}} \,\ , 
\end{eqnarray}
where
\begin{eqnarray} 
{\cal L}_{SUSY} &=&   {\cal L}_{\mbox{Lepton}}+ {\cal L}_{\mbox{Quarks}}+ 
{\cal L}_{\mbox{Gauge}}+ {\cal L}_{\mbox{Scalar}}, 
\end{eqnarray}
is the supersymmetric part while ${\cal L}_{\mbox{soft}}$ break superssymetry.

Where we define
\begin{eqnarray} 
{\cal L}_{Lepton}&=& \int d^{4}\theta\;\left[\,
\hat{ \bar{L}}e^{2g\hat{V}} \hat{L} \,\right],
\label{lepsusy}
\end{eqnarray}
this term give us the following interactions
\begin{eqnarray}
{\cal L}_{F}&=& \vert F_{L} \vert^2; \,\
{\cal L}_{D}= \frac{g}{2} \bar{\tilde{L}} \left( \lambda^{a}D^{a} \right)\tilde{L} , \nonumber \\
{\cal L}_{llV}&=&\frac{g}{2} \left[ \bar{L}\bar{\sigma}^{m} \left( \lambda^{a}V^{a}_{m} \right) L \right] ; \,\
{\cal L}_{ \tilde{l} \tilde{l}V}=- \frac{\imath g}{2}\left[ 
\tilde{L}\left( \lambda^{a}V^{a}_{m} \right) \partial^{m}\bar{\tilde{L}}- 
\bar{\tilde{L}}\left( \lambda^{a}V^{a}_{m} \right) \partial^{m} \tilde{L} \right], \nonumber \\
{\cal L}_{l \tilde{l} \tilde{V}}&=&- \frac{\imath g}{ \sqrt{2}}
\left[ \bar{L}\left( \lambda^{a}\bar{\lambda}^{a}_{A} \right)\tilde{L}- 
\bar{\tilde{L}}\left( \lambda^{a}\lambda^{a}_{A} \right) L \right]; \,\
{\cal L}_{ \tilde{l} \tilde{l}VV}= \frac{g^2}{4} \left[
\bar{\tilde{L}} \left( \lambda^{a}V^{a}_{m}\right) \left( \lambda^{b}V^{bm}\right) \tilde{L} \right] , \nonumber \\
\label{lepbos}
\end{eqnarray}
where
\begin{eqnarray}
\hat{V}&=& \left( T^{a}\hat{V}^{a} \right) = \left( \begin{array}{ccc} 
V^{3}_{m}+ \frac{1}{\sqrt{3}}V^{8}_{m} & \sqrt{2} W^{+}_{m} & \sqrt{2} V^{-}_{m} \\ 
\sqrt{2} W^{-}_{m} &- V^{3}_{m}+ \frac{1}{\sqrt{3}}V^{8}_{m} & \sqrt{2} U^{--}_{m} \\
\sqrt{2} V^{+}_{m} & \sqrt{2} U^{++}_{m} &- \frac{2}{\sqrt{3}}V^{8}_{m}         
\end{array} \right),  
\label{TVmatriz}
\end{eqnarray} 
Using the first relation in Eq.(\ref{lepbos}) together with Eq.(\ref{TVmatriz}) we get
\begin{eqnarray}
{\cal L}_{llV}&=&\left[ \bar{\nu}_{a}\bar{\sigma}^{m}\nu_{a} \left( V^{3}_{m}+ \frac{V^{8}_{m}}{\sqrt{3}} \right) -
\bar{l}_{a}\bar{\sigma}^{m}l_{a} \left( V^{3}_{m}- \frac{V^{8}_{m}}{\sqrt{3}} \right) -
\bar{l}^{c}_{a}\bar{\sigma}^{m}l^{c}_{a} \left( \frac{2V^{8}_{m}}{\sqrt{3}} \right) 
\right. \nonumber \\ &-& \left.
\sqrt{2}\bar{\nu}_{a}\bar{\sigma}^{m}l_{a}W^{+}_{m}- \sqrt{2}\bar{l}_{a}\bar{\sigma}^{m}\nu_{a}W^{-}_{m}- 
\sqrt{2}\bar{\nu}_{a}\bar{\sigma}^{m}l^{c}_{a}V^{-}_{m}- \sqrt{2}\bar{l}^{c}_{a}\bar{\sigma}^{m}\nu_{a}V^{+}_{m}
\right. \nonumber \\ &-& \left.
\sqrt{2}\bar{l}_{a}\bar{\sigma}^{m}l^{c}_{a}U^{--}_{m}- \sqrt{2}\bar{l}^{c}_{a}\bar{\sigma}^{m}l_{a}U^{++}_{m}
\right],
\label{lepbos1}
\end{eqnarray} 
We will define the following Dirac spinors for the leptons
\begin{eqnarray}
\Psi(l)&=& \left( \begin{array}{c}
             l \\
	     \bar{l}^{c} 
\end{array} \right), \,\
\Psi^c(l)= \left( \begin{array}{c}
             l^c \\
	     \bar{l} 
\end{array} \right), \,\
\Psi( \nu_{l})= \left( \begin{array}{c}
             \nu_{l} \\
	     0 
\end{array} \right).
\label{defspoinorsleptons}
\end{eqnarray}

We can rewrite Eq.(\ref{lepbos1}) as follows
\begin{eqnarray}
{\cal L}_{llV}&=&- \frac{g}{ \sqrt{2}} \left( \bar{ \Psi^{c}}(l) 
\gamma^{m}L \Psi(l)U^{++}_{m}+ \bar{ \Psi^{c}}(l)
\gamma^{m}L \Psi( \nu_{l})V^{+}_{m}+ \bar{ \Psi}( \nu_{l}) 
\gamma^{m}L \Psi(l)W^{+}_{m}+hc \right) \nonumber \\
&-& \frac{gt}{ \sqrt{1+4t^{2}}} \bar{ \Psi}(l) \gamma^{m} \Psi(l)A_{m}
- \frac{g}{2} \sqrt{ \frac{1+4t^{2}}{1+3t^{2}}} 
\bar{ \Psi}( \nu_{l}) \gamma^{m}L \Psi( \nu_{l}) \left[ Z_{m}-
\frac{1}{ \sqrt{3}} \frac{1}{ \sqrt{1+4t^{2}}} Z^{ \prime}_{m} \right]
\nonumber \\
&-& \frac{g}{4} \sqrt{ \frac{1+4t^{2}}{1+3t^{2}}} 
\left[ \left( \frac{-1}{1+4t^{2}} \bar{ \Psi}(l) \gamma^{m} \Psi(l)
+ \bar{ \Psi}(l) \gamma^{m} \gamma^{5} \Psi(l) \right)Z_{m} \right. \nonumber\\
&+& \left. \left(
\frac{- \sqrt{3}}{ \sqrt{1+4t^{2}}} \bar{ \Psi}(l) \gamma^{m} \Psi(l)-
\frac{1}{ \sqrt{3}} \frac{1}{ \sqrt{1+4t^{2}}} 
\bar{ \Psi}(l) \gamma^{m} \gamma^{5} \Psi(l) \right) Z^{ \prime}_{m} \right] 
\nonumber \\
&=& {\cal L}^{NC}_{llV}+ {\cal L}^{CC}_{llV},
\label{lepbosint}
\end{eqnarray}
where $t$ is defined in Eq.(\ref{tinthetawm331}).

Let us define also the following parameters
\begin{eqnarray}
e&=& \frac{g \sin \theta}{ \sqrt{1+3 \sin^{2} \theta}}, \nonumber \\
h(t)&=&1+4t^{2}, \nonumber \\
v_{l}&=&- \frac{1}{h(t)}, \nonumber \\
a_{l}&=&1 ,\nonumber \\
v^{ \prime}_{l}&=&- \frac{ \sqrt{3}}{ \sqrt{h(t)}}, \nonumber \\
a^{ \prime}_{l}&=& \frac{v^{ \prime}_{l}}{3}.
\label{pardef}
\end{eqnarray}
Then we can now rewrite the neutral part of 
Eq.(\ref{lepbosint}):
\begin{eqnarray}
{\cal L}^{NC}_{llV}&=&-e \bar{ \Psi}(l) \gamma^{m} \Psi(l)A_{m}-
\frac{g}{2} \frac{M_{Z}}{M_{W}}
\bar{ \Psi}( \nu_{l}) \gamma^{m}L \Psi( \nu_{l}) \left[ Z_{m}-
\frac{1}{ \sqrt{3}} \frac{1}{ \sqrt{h(t)}} Z^{ \prime}_{m} \right] \nonumber \\
&-& \frac{g}{4} \frac{M_{Z}}{M_{W}} \left[ \bar{ \Psi}(l) 
\gamma^{m}
(v_{l}+a_{l} \gamma^{5}) \Psi(l) Z_{m}+ \bar{ \Psi}(l) \gamma^{m}
(v^{ \prime}_{l}+a^{ \prime}_{l} \gamma^{5}) \Psi(l) Z^{ \prime}_{m} \right],
\label{lagr1}
 \end{eqnarray}
The charged part of 
Eq.(\ref{lepbosint}) is
\begin{eqnarray}
{\cal L}^{CC}_{llV}&=&- \frac{g}{ \sqrt{2}} \left( \bar{ \Psi^{c}}(l) 
\gamma^{m}L \Psi(l)U^{++}_{m}+ \bar{ \Psi^{c}}(l)
\gamma^{m}L \Psi( \nu_{l})V^{+}_{m}+ \bar{ \Psi}( \nu_{l}) 
\gamma^{m}L \Psi(l)W^{+}_{m}+hc \right) ,
\label{lagr2}
\end{eqnarray}
with $g$ such that
\begin{equation}
g^2=\frac{8G_F M^2_W}{\sqrt2};\quad {\rm or}\quad 
g^2/\alpha=4\pi\sin^2\theta_W,
\label{gf}
\end{equation}
with $\alpha\approx 1/127$. The lagrangian in the Eqs.(\ref{lagr1},\ref{lagr2}) is the same that it was shown in 
\cite{ppf,dumm2}.

Now, we will consider the quarks
 \begin{eqnarray} 
{\cal L}_{Quarks}&=& \int d^{4}\theta\;\left[\,\hat{ \bar{Q}}_{1}
e^{2[g(\hat{V}_{c}+\hat{V})+(2g'/3)\hat{V}']} \hat{Q}_{1} 
+\,\hat{\bar{Q}}_{\alpha}
e^{2[g(\hat{V}_{c}+\hat{V})-(g'/3)\hat{V}']} \hat{Q}_{\alpha} \,\right. 
\nonumber \\&+& 
\left.\,\hat{ \bar{u}}_{i}
e^{2[g(\hat{V}_{c}+\hat{V})-(2g'/3)\hat{V}']} \hat{u}_{i} 
+\hat{ \bar{d}}_{i}
e^{2[g(\hat{V}_{c}+\hat{V})+(g'/3)\hat{V}']} \hat{d}_{i}\right.
\nonumber \\ &+&\left. \,\hat{ \bar{J}}
e^{2[g(\hat{V}_{c}+\hat{V}) -(5g'/3)\hat{V}']} \hat{J} 
+ \hat{ \bar{j}}_i
e^{2[g(\hat{V}_{c}+\hat{V}) +(4g'/3)\hat{V}']} \hat{j}_i\right], 
\label{quarkbos}
\end{eqnarray}
and in similar way as we have done for the leptons we have the followings interactions
\begin{eqnarray}
{\cal L}_{F}&=& \vert F_{Q_i} \vert^2 +
 \vert F_{u_i} \vert^2 + \vert F_{d_i} \vert^2 + 
\vert F_{J_i} \vert^2, \nonumber \\
{\cal L}_{D}&=& \frac{g_{s}}{2}( \bar{\tilde{Q}}_{i}\lambda^a\tilde{Q}_{i} -
\bar{\tilde{u}}^{c}_{i}\lambda^{*a}\tilde{u}^{c}_{i}- 
\bar{\tilde{d}}^{c}_{i}\lambda^{*a}\tilde{d}^{c}_{i}-
\bar{\tilde{J}}^{c}_{i}\lambda^{*a}\tilde{J}^{c}_{i})D^a_c 
+\frac{g}{2} \left( \bar{\tilde{Q}}_{3}\lambda^a\tilde{Q}_{3}-
\bar{\tilde{Q}}_{ \alpha}\lambda^{*a}\tilde{Q}_{ \alpha}  \right) D^{a} \nonumber \\
&+& \frac{g^{\prime}}{2} \left[ \frac{2}{3} \bar{\tilde{Q}}_{3}\tilde{Q}_{3} -
 \frac{1}{3} \bar{\tilde{Q}}_{ \alpha}\tilde{Q}_{ \alpha}
- \frac{2}{3} \bar{\tilde{u}}^{c}_{i}\tilde{u}^{c}_{i}+ 
\frac{1}{3} \bar{\tilde{d}}^{c}_{i}\tilde{d}^{c}_{i}-
\frac{5}{3} \bar{\tilde{J}}^{c}\tilde{J}^{c}+
\frac{4}{3} \bar{\tilde{j}}^{c}_{ \beta}\tilde{j}^{c}_{ \beta}
 \right] D, \nonumber \\
{\cal L}_{qqV}&=& \frac{g_{s}}{2} ( 
\bar{Q}_{i}\bar\sigma^m\lambda^a Q_{i}- 
\bar{u}^{c}_i\bar\sigma^m\lambda^{*a} u^{c}_i-
\bar{d}^{c}_i\bar\sigma^m\lambda^{*a} d^{c}_i-
\bar{J}^{c}_i\bar\sigma^m\lambda^{*a} J^{c}_i)g^a_m  \nonumber \\ 
&+& \frac{g}{2}( \bar{Q}_{3}\bar\sigma^m\lambda^a Q_{3}- 
\bar{Q}_{ \alpha}\bar\sigma^m\lambda^{*a} Q_{ \alpha})V^a_m  \nonumber \\
&+& \frac{g^{ \prime}}{2} \left( 
\frac{2}{3} \bar{Q}_{3}\bar\sigma^m Q_{3}- 
\frac{1}{3} \bar{Q}_{ \alpha}\bar\sigma^m Q_{ \alpha}-
\frac{2}{3} \bar{u}^{c}_i\bar\sigma^m u^{c}_i+
\frac{1}{3} \bar{d}^{c}_i\bar\sigma^m d^{c}_i-
\frac{5}{3} \bar{J}^{c}\bar\sigma^m J^{c}+
\frac{4}{3} \bar{j}^{c}_{ \beta}\bar\sigma^m j^{c}_{ \beta} 
\right)V_m, \nonumber \\
{\cal L}_{ \tilde{q} \tilde{q}V}&=& \frac{-ig_{s}}{2} \left[ (
\tilde{Q}_{i}\lambda^a\partial^m\bar{\tilde{Q}}_{i}-
\bar{\tilde{Q}}_{i}\lambda^a\partial^m \tilde{Q}_{i}-
\tilde{u}^{c}_i\lambda^{*a}\partial^m\bar{\tilde{u}}^{c}_i + 
\bar{\tilde{u}}^{c}_i\lambda^{*a}\partial^m \tilde{u}^{c}_i \right. \nonumber \\
&-& \left.
\tilde{d}^{c}_i\lambda^{*a}\partial^m\bar{\tilde{d}}^{c}_i+
\bar{\tilde{d}}^{c}_i\lambda^{*a}\partial^m \tilde{d}^{c}_i -
\tilde{J}^{c}_i\lambda^{*a}\partial^m\bar{\tilde{J}}^{c}_i+
\bar{\tilde{J}}^{c}_i\lambda^{*a}\partial^m \tilde{J}^{c}_i)g^a_m \right] \nonumber \\ 
&-& \frac{ig}{2} ( \tilde{Q}_{3}\lambda^a\partial^m\bar{\tilde{Q}}_{3}-
\bar{\tilde{Q}}_{3}\lambda^a\partial^m \tilde{Q}_{3}-
\tilde{Q}_{ \alpha}\lambda^{*a}\partial^m\bar{\tilde{Q}}_{ \alpha}+
\bar{\tilde{Q}}_{ \alpha}\lambda^{*a}\partial^m \tilde{Q}_{ \alpha})V^a_m \nonumber \\
&-& \frac{ig^{ \prime}}{2} \left[ \frac{2}{3} (
\tilde{Q}_{3} \partial^m\bar{\tilde{Q}}_{3}-
\bar{\tilde{Q}}_{3} \partial^m \tilde{Q}_{3})- \frac{1}{3} (
\tilde{Q}_{ \alpha} \partial^m\bar{\tilde{Q}}_{ \alpha}-
\bar{\tilde{Q}}_{ \alpha} \partial^m \tilde{Q}_{ \alpha})-
\frac{2}{3} ( \tilde{u}^{c}_i \partial^m\bar{\tilde{u}}^{c}_i-
\bar{\tilde{u}}^{c}_i \partial^m \tilde{u}^{c}_i) \right. \nonumber \\
&+& \left. \frac{1}{3}
(\tilde{d}^{c}_i \partial^m\bar{\tilde{d}}^{c}_i-
\bar{\tilde{d}}^{c}_i \partial^m \tilde{d}^{c}_i)- \frac{5}{3}
(\tilde{J}^{c} \partial^m\bar{\tilde{J}}^{c}-
\bar{\tilde{J}}^{c} \partial^m \tilde{J}^{c})+ \frac{4}{3}
(\tilde{j}^{c}_{ \beta} \partial^m\bar{\tilde{j}}^{c}_{ \beta}-
\bar{\tilde{j}}^{c}_{ \beta} \partial^m \tilde{j}^{c}_{ \beta}) \right] V_m, \nonumber \\
{\cal L}_{q \tilde{q} \tilde{V}}&=& \frac{-ig_{s}}{ \sqrt{2}} \left[ ( 
\bar{Q}_{i}\lambda^a\tilde{Q}_{i}-
\bar{u}^{c}_i\lambda^{*a}\tilde{u}^{c}_i- \bar{d}^{c}_i\lambda^{*a}\tilde{d}^{c}_i-
\bar{J}^{c}_i\lambda^{*a}\tilde{J}^{c}_i) \bar{\lambda}^a_{c} \right. \nonumber \\
&-& \left. ( \bar{\tilde{Q}}_{i}\lambda^aQ_{i}-
\bar{\tilde{u}}^{c}_i\lambda^{*a}u^{c}_i-
\bar{\tilde{d}}^{c}_i\lambda^{*a}d^{c}_i- \bar{\tilde{J}}^{c}_i\lambda^{*a}J^{c}_i)
\lambda^a_{c} \right] \nonumber \\
&-& \frac{ig}{ \sqrt{2}} \left[ ( 
\bar{Q}_{3}\lambda^a\tilde{Q}_{3}- \bar{Q}_{\alpha}\lambda^{*a}\tilde{Q}_{\alpha})\bar{\lambda}^a_{A}- ( \bar{\tilde{Q}}_{3}\lambda^aQ_{3}-
\bar{\tilde{Q}}_{\alpha}\lambda^{*a}Q_{\alpha}) \lambda^a_{A} \right] \nonumber \\
&-& \frac{ig'}{\sqrt{2}} \left[ \left( \frac{2}{3} \bar{Q}_{3}\tilde{Q}_{3}- 
\frac{1}{3}\bar{Q}_{\alpha}\tilde{Q}_{\alpha}-\frac{2}{3} \bar{u}^{c}_i\tilde{u}^{c}_i+
\frac{1}{3} \bar{d}^{c}_i\tilde{d}^{c}_i-\frac{5}{3} \bar{J}^{c}\tilde{J}^{c}+
\frac{4}{3} \bar{j}^{c}_{ \beta}\tilde{j}^{c}_{ \beta} \right) \bar{\lambda}_{B} 
\right. \nonumber \\
&-& \left. \left( 
\frac{2}{3}\bar{\tilde{Q}}_{3}Q_{3}- \frac{1}{3}\bar{\tilde{Q}}_{\alpha}Q_{\alpha}-
\frac{2}{3} \bar{\tilde{u}}^{c}_iu^{c}_i+\frac{1}{3} \bar{\tilde{d}}^{c}_id^{c}_i-
\frac{5}{3} \bar{\tilde{J}}^{c}J^{c}+
\frac{4}{3} \bar{\tilde{j}}^{c}_{ \beta}j^{c}_{ \beta} \right) \lambda_B \right], 
\nonumber \\
{\cal L}_{ \tilde{q} \tilde{q}VV}&=& \frac{-1}{4} \left[ g_{s}^2(
\bar{\tilde{Q}}_{i}\lambda^{a}\lambda^{b} \tilde{Q}_{i}+
\bar{\tilde{u}}^{c}_i\lambda^{*a}\lambda^{*b} \tilde{u}^{c}_i+
\bar{\tilde{d}}^{c}_i\lambda^{*a}\lambda^{*b} \tilde{d}^{c}_i+
\bar{\tilde{J}}^{c}_i\lambda^{*a}\lambda^{*b} \tilde{J}^{c}_i)g^a_mg^{bm} 
\right] \nonumber \\
&-& \frac{1}{4} \left[ g^2(
\bar{\tilde{Q}}_{3}\lambda^{a}\lambda^{b} \tilde{Q}_{3}+
\bar{\tilde{Q}}_{ \alpha}\lambda^{*a}\lambda^{*b} \tilde{Q}_{ \alpha}) 
\right] V^a_mV^{bm} 
- \frac{1}{2} \left[ g_{s}g(
\bar{\tilde{Q}}_{3}\lambda^{a}\lambda^{b} \tilde{Q}_{3}+
\bar{\tilde{Q}}_{ \alpha}\lambda^{a}\lambda^{*b} \tilde{Q}_{ \alpha}) 
\right] g^a_mV^{bm} \nonumber \\
&-&  \frac{1}{2}g_{s}g^{ \prime} \left[ 
\frac{2}{3} \bar{\tilde{Q}}_{3}\lambda^a \tilde{Q}_{3}-
\frac{1}{3} \bar{\tilde{Q}}_{ \alpha}\lambda^a \tilde{Q}_{ \alpha}+
\frac{2}{3} \bar{\tilde{u}}^{c}_i\lambda^{*a} \tilde{u}^{c}_i-
\frac{1}{3} \bar{\tilde{d}}^{c}_i\lambda^{*a} \tilde{d}^{c}_i+
\frac{5}{3} \bar{\tilde{J}}^{c}\lambda^{*a} \tilde{J}^{c} \right. \nonumber \\
&-& \left.
\frac{4}{3} \bar{\tilde{j}}^{c}_{ \beta}\lambda^{*a} \tilde{j}^{c}_{ \beta} \right] 
g^{am}V_m-  \frac{1}{2}gg^{ \prime} \left[ 
\frac{2}{3} \bar{\tilde{Q}}_{3}\lambda^a \tilde{Q}_{3}+
\frac{1}{3} \bar{\tilde{Q}}_{ \alpha}\lambda^{*a} \tilde{Q}_{ \alpha} \right] 
V^{am}V_m \nonumber \\
&-& \frac{1}{4} g^{ \prime 2} \left[ 
\frac{4}{9}(\bar{\tilde{Q}}_{3}\tilde{Q}_{3}+\bar{\tilde{u}}^{c}_i\tilde{u}^{c}_i)+
\frac{1}{9}(\bar{\tilde{Q}}_{\alpha}\tilde{Q}_{\alpha}+
\bar{\tilde{d}}^{c}_i\tilde{d}^{c}_i)+ \frac{25}{9} \bar{\tilde{J}}^{c}\tilde{J}^{c}+
\frac{16}{9} \bar{\tilde{j}}^{c}_{\beta}\tilde{j}^{c}_{\beta} \right] V^m V_m. \nonumber \\
\label{quarksbos}
\end{eqnarray}
where
\begin{eqnarray}
\hat{\bar{V}}&=&\left( \bar{T}^{a}\hat{V}^{a} \right) = \left( \begin{array}{ccc} 
V^{3}_{m}+ \frac{1}{\sqrt{3}}V^{8}_{m} & \sqrt{2} W^{+}_{m} & \sqrt{2} V^{+}_{m} \\ 
\sqrt{2} W^{-}_{m} &- V^{3}_{m}+ \frac{1}{\sqrt{3}}V^{8}_{m} & \sqrt{2} U^{++}_{m} \\
\sqrt{2} V^{-}_{m} & \sqrt{2} U^{--}_{m} &- \frac{2}{\sqrt{3}}V^{8}_{m}         
\end{array} \right).
\label{barTVmatriz}
\end{eqnarray} 
as we have done with the leptons we can write
\begin{eqnarray}
{\cal L}^{Q_{\alpha}}_{qqV}&=&\left[ 
\bar{d}_{\alpha}\bar{\sigma}^{m}d_{\alpha} \left( V^{3}_{m}+ \frac{V^{8}_{m}}{\sqrt{3}} \right) -
\bar{u}_{\alpha}\bar{\sigma}^{m}u_{\alpha} \left( V^{3}_{m}- \frac{V^{8}_{m}}{\sqrt{3}} \right) -
\bar{j}_{\alpha}\bar{\sigma}^{m}j_{\alpha} \left( \frac{2V^{8}_{m}}{\sqrt{3}} \right) 
\right. \nonumber \\ &-& \left.
\sqrt{2}\bar{u}_{\alpha}\bar{\sigma}^{m}d_{\alpha}W^{+}_{m}- \sqrt{2}\bar{d}_{\alpha}\bar{\sigma}^{m}u_{\alpha}W^{-}_{m}- 
\sqrt{2}\bar{j}_{\alpha}\bar{\sigma}^{m}d_{\alpha}V^{-}_{m}- \sqrt{2}\bar{d}_{\alpha}\bar{\sigma}^{m}j_{\alpha}V^{+}_{m}
\right. \nonumber \\ &-& \left.
\sqrt{2}\bar{j}_{\alpha}\bar{\sigma}^{m}u_{\alpha}U^{--}_{m}- \sqrt{2}\bar{u}_{\alpha}\bar{\sigma}^{m}j_{\alpha}U^{++}_{m}
\right], \nonumber \\
{\cal L}^{Q_{3}}_{qqV}&=&\left[ 
\bar{u}_{3}\bar{\sigma}^{m}u_{3} \left( V^{3}_{m}+ \frac{V^{8}_{m}}{\sqrt{3}} \right) -
\bar{d}_{3}\bar{\sigma}^{m}d_{3} \left( V^{3}_{m}- \frac{V^{8}_{m}}{\sqrt{3}} \right) -
\bar{J}\bar{\sigma}^{m}J \left( \frac{2V^{8}_{m}}{\sqrt{3}} \right) 
\right. \nonumber \\ &-& \left.
\sqrt{2}\bar{u}_{3}\bar{\sigma}^{m}d_{3}W^{+}_{m}- \sqrt{2}\bar{d}_{3}\bar{\sigma}^{m}u_{3}W^{-}_{m}- 
\sqrt{2}\bar{u}_{3}\bar{\sigma}^{m}JV^{-}_{m}- \sqrt{2}\bar{J}\bar{\sigma}^{m}u_{3}V^{+}_{m}
\right. \nonumber \\ &-& \left.
\sqrt{2}\bar{d}_{3}\bar{\sigma}^{m}JU^{--}_{m}- \sqrt{2}\bar{J}\bar{\sigma}^{m}u_{3}U^{++}_{m}
\right],
\end{eqnarray} 
We will define the following Dirac spinors for the leptons
\begin{eqnarray}
\Psi(u)&=& \left( \begin{array}{c}
             u \\
	\bar{u}^{c} 
\end{array} \right), \,\
\Psi^c(u)= \left( \begin{array}{c}
             u^c \\
	     \bar{u} 
\end{array} \right), \nonumber \\
\Psi(d)&=& \left( \begin{array}{c}
             d \\
	\bar{d}^{c} 
\end{array} \right), \,\
\Psi^c(d)= \left( \begin{array}{c}
             d^c \\
	     \bar{d} 
\end{array} \right), \nonumber \\
\Psi(J)&=& \left( \begin{array}{c}
             J \\
	\bar{J}^{c} 
\end{array} \right), \,\
\Psi^c(J)= \left( \begin{array}{c}
             J^c \\
	     \bar{J} 
\end{array} \right), \nonumber \\
\Psi(j)&=& \left( \begin{array}{c}
             j \\
	\bar{j}^{c} 
\end{array} \right), \,\
\Psi^c(j)= \left( \begin{array}{c}
             j^c \\
	     \bar{j} 
\end{array} \right).
\label{defspoinorsleptons}
\end{eqnarray}
The first lagrangian in Eq.(\ref{lepbos}) and Eq.(\ref{quarksbos})  give us the lagrangian presented in Eqs.(\ref{lf},\ref{lq}) and 
therefore the phenomenological studies presented in \cite{Pleitez:2021abk} are still hold in this model. 

The kinectic terms and self-interaction for the gauge bosons are
\begin{eqnarray}
{\cal L}_{Gauge}&=& \frac{1}{4} \left[ \int  d^{2}\theta\; 
[ {W}^{a}_{c}{W}^{a}_{c}+{W}^{a}{W}^{a}+{W}^{ \prime}{W}^{ \prime}
\int  d^{2}\bar{\theta}\; 
[ \bar{W}^{a}_{c}\bar{W}^{a}_{c}+\bar{W}^{a}\bar{W}^{a}+
\bar{W}^{ \prime}\bar{W}^{ \prime}]\,\right],
\label{l5}
\end{eqnarray}
where $\hat{V}_{c}=T^{a}\hat{V}^{a}_{c}$, 
$\hat{V}=T^{a}\hat{V}^{a}$ and $T^a=\lambda^{a}/2$ are the generators of 
$SU(3)$ i.e., $a=1,\cdots,8$, and $g$ and $g^{\prime}$ are the gauge 
coupling of $SU(3)_L$ and $U(1)_N$. $W^{a}_{c}$, $W^{a}$ and $W^{ \prime}$ 
are the strength fields, and they are given by 

\begin{eqnarray}
W^{a}_{\alpha c}&=&- \frac{1}{8g} \bar{D} \bar{D} e^{-2g \hat{V}_{c}} 
D_{\alpha} e^{-2g \hat{V}_{c}} \nonumber \\
W^{a}_{\alpha}&=&- \frac{1}{8g} \bar{D} \bar{D} e^{-2g \hat{V}} 
D_{\alpha} e^{-2g \hat{V}} \nonumber \\
W^{\prime}_{\alpha}&=&- 
\frac{1}{4} \bar{D} \bar{D} D_{\alpha} \hat{V}^{\prime} \,\ .
\label{l6}
\end{eqnarray}

Finally 
\begin{eqnarray}
  {\cal L}_{Scalar} 
     &=&  \int d^{4}\theta\;\left[\,\hat{ \bar{ \eta}}e^{2g\hat{V}}   
                         \hat{ \eta} +			  
\hat{ \bar{ \rho}}e^{2g\hat{V}+g'\hat{V}'}                            
\hat{ \rho} +
\hat{ \bar{ \chi}}e^{2g\hat{V}-g'\hat{V}'}                            
\hat{ \chi} + 
\hat{ \bar{S}}e^{2g\hat{V}}                            
\hat{S}\right.
\nonumber \\ &+&
\left. \hat{ \bar{ \eta}}^\prime e^{2g\hat{\bar V}}   
                         \hat{ \eta}^\prime +			  
\hat{\bar{ \rho}}^\prime e^{2g\hat{\bar V}-g'\hat{\bar V}'}                     
\hat{ \rho}^\prime +
\hat{ \bar{ \chi}}^\prime e^{2g\hat{\bar V}+g'\hat{\bar V}'}                    
\hat{ \chi}^\prime + 
\hat{ \bar{S}}^\prime e^{2g\hat{\bar V}}                            
\hat{S}^\prime\right]\nonumber \\ &+&
\int d^2\theta\, W+\int d^2\bar\theta\, \bar{W},
\label{l7}
\end{eqnarray}
where $W$ is the superpotential. 

The interaction of the fermions with the usual scalars are
\begin{eqnarray}
{\cal L}_{F}&=& \vert F_{\eta} \vert^2+ \vert F_{\rho} \vert^2+ 
\vert F_{\chi} \vert^2+ \vert F_{S} \vert^2 + 
\vert F_{\eta^{\prime}} \vert^2+ 
\vert F_{\rho^{\prime}} \vert^2+ 
\vert F_{\chi^{\prime}} \vert^2+ \vert F_{S^{\prime}} \vert^2, \nonumber \\
{\cal L}_{D}&=& \frac{g}{2} \left[ \bar{\eta}\lambda^a\eta+ 
\bar{\rho}\lambda^a\rho+ \bar{\chi}\lambda^a\chi+ 
\bar{S}\lambda^aS- 
\bar{\eta}^{\prime}\lambda^{* a}\eta^{\prime}- 
\bar{\rho}^{\prime}\lambda^{* a}\rho^{\prime}- 
\bar{\chi}^{\prime}\lambda^{* a}\chi^{\prime}- 
\bar{S}^{\prime}\lambda^{* a}S^{\prime} \right] D^{a} \nonumber \\
&+& \frac{g^{ \prime}}{2} \left[  
\bar{\rho}\rho- \bar{\chi}\chi- \bar{\rho}^{\prime}\rho^{\prime}+ 
\bar{\chi}^{\prime}\chi^{\prime} \right]D, \nonumber \\
{\cal L}_{llH}&=&- 
\left[ \lambda^{ij}_{2} \left( \epsilon LL\eta \right) + \lambda^{ij}_{3} \left( LLS \right) \right] \nonumber \\
&=&\left\{ \lambda^{ij}_{2} \left[ 
\left( l_{i}l^{c}_{j}- l^{c}_{i}l_{j} \right) \eta^{0}+
\left( l^{c}_{i}\nu_{j}- \nu_{i}l^{c}_{j} \right) \eta^{-}_{1}+
\left( \nu_{i}l_{j}-l_{i} \nu_{j} \right) \eta^{+}_{2} \right] + \lambda^{ij}_{3}\left[ \nu_{i}\nu_{j}\sigma^{0}_{1} 
\right. \right. \nonumber \\ &+& \left. \left.
\left( l_{i}l^{c}_{j}+l^{c}_{i}l_{j}\right) \sigma^{0}_{2}+
\left( \nu_{i}l^{c}_{j}+l^{c}_{i}\nu_{j} \right) h^{-}_{1}+
\left( \nu_{i}l_{j}+l_{i}\nu_{j} \right) h^{+}_{2}+ l_{i}l_{j}H^{++}_{1}+  l^{c}_{i}l^{c}_{j}H^{--}_{2} 
\right]
\right\}, \nonumber \\
{\cal L}_{qqH}&=&- \left[ 
\kappa_{1} Q_3\eta^{\prime}u^{c}_{i}+
\kappa_{2} Q_3\rho^{\prime}d^{c}_{i}+
\kappa_{3} Q_3\chi^{\prime}J^{c}+
\kappa_{4} Q_{\alpha}\eta d^{c}_{i}+
\kappa_{5} Q_{\alpha}\rho u^{c}_{i}+
\kappa_{6} Q_{\alpha}\chi j^{c}_{\beta} \right] \nonumber \\
&=&- \left[
\kappa_{1} \left( u_{3}u^{c}_{i}\eta^{\prime 0}+d_{3}u^{c}_{i}\eta^{\prime +}_{1}+Ju^{c}_{i}\eta^{\prime -}_{2}\right) +
\kappa_{2} \left( u_{3}d^{c}_{i}\rho^{\prime -}+d_{3}d^{c}_{i}\rho^{\prime 0}+Jd^{c}_{i}\rho^{\prime --}\right) 
\right. \nonumber \\ &+& \left.
\kappa_{3} \left( u_{3}J^{c}\chi^{\prime +}+d_{3}J^{c}\chi^{\prime ++}+JJ^{c}\chi^{\prime 0}\right) +
\kappa_{4} \left( d_{\alpha}d^{c}_{i}\eta^{0}+u_{\alpha}d^{c}_{i}\eta^{-}_{1}+j_{\alpha}d^{c}_{i}\eta^{+}_{2}\right)
\right. \nonumber \\ &+& \left.
\kappa_{5} \left( d_{\alpha}u^{c}_{i}\rho^{+}+u_{\alpha}u^{c}_{i}\rho^{0}+j_{\alpha}u^{c}_{i}\rho^{++}\right) +
\kappa_{6} \left(d_{\alpha}j^{c}_{\beta}\chi^{-}+u_{\alpha}j^{c}_{\beta}\chi^{--}+j_{\alpha}j^{c}_{\beta}\chi^{ 0}\right)
\right].
\label{ffH}
\end{eqnarray}

The superpotential of the model is
\begin{equation}
W=W_{2}+W_{3},
\end{equation}
where
The superpotencial in MSUSY331 is defined as \cite{331susy1,mcr,Rodriguez:2005jt}
\begin{equation}
W=W_{2}+ W_{3}+hc, 
\label{sp1}
\end{equation}
where each term above is defined as follows
\begin{eqnarray}
W_{2}&=&\mu_{0i}(\hat{L}_{i}\hat{\eta}^{\prime})+ 
\mu_{ \eta} (\hat{\eta}\hat{\eta}^{\prime})+
\mu_{ \rho} (\hat{\rho}\hat{\rho}^{\prime})+ 
\mu_{ \chi} (\hat{\chi}\hat{\chi}^{\prime})+
\mu_{S} Tr[(\hat{S} \hat{S}^{\prime})]. 
\label{w2geral}
\end{eqnarray} 
All the coefficients in the equation above have mass dimension. The last term of the 
superpotential in MSUSY331 has the following form
\begin{eqnarray}
W_{3}&=&\lambda_{1ijk} 
(\epsilon \hat{L}_{i}\hat{L}_{j}\hat{L}_{k})+
\lambda_{2ij} (\epsilon \hat{L}_{i}\hat{L}_{j}\hat{\eta})+
\lambda_{3ij} 
(\hat{L}_{i}\hat{S}\hat{L}_{j}) +
\lambda_{4i}(\epsilon\hat{L}_{i}\hat{\chi}\hat{\rho})  \nonumber \\
&+&
f_{1} (\epsilon \hat{\rho} \hat{\chi}\hat{\eta})+
f_{2} (\hat{\chi}\hat{S}\hat{\rho})  +
f_{3} (\hat{\eta}\hat{S}\hat{\eta}) +
f_{4}\epsilon_{ijk}\epsilon_{lmn}\hat{S}_{il}\hat{S}_{jm}\hat{S}_{kn}+
f^{\prime}_{1} (\epsilon \hat{\rho}^{\prime}\hat{\chi}^{\prime} \hat{\eta}^{\prime})+
f^{\prime}_{2} (\hat{\chi}^{\prime} 
\hat{S}^{\prime}\hat{\rho}^{\prime})  
\nonumber \\ &+&
f^{\prime}_{3} (\hat{\eta}^{\prime} 
\hat{S}^{\prime}\hat{\eta}^{\prime}) +
f^{\prime}_{4}\epsilon_{ijk}\epsilon_{lmn}
\hat{S}^{\prime}_{il}\hat{S}^{\prime}_{jm}
\hat{S}^{\prime}_{kn} +
\kappa_{1i} (\hat{Q}_{3} \hat{\eta}^{\prime}) \hat{u}^{c}_{i}+ 
\kappa_{2i} (\hat{Q}_{3} \hat{\rho}^{\prime}) \hat{d}^{c}_{i} +
\kappa_{3} 
(\hat{Q}_{3}\hat{\chi}^{\prime}) 
\hat{J}^{c} 
\nonumber \\
&+&
\kappa_{4\alpha i} (\hat{Q}_{\alpha} \hat{\eta}) \hat{d}^{c}_{i}+
\kappa_{5\alpha i} (\hat{Q}_{\alpha} \hat{\rho}) \hat{u}^{c}_{i} +
\kappa_{6\alpha\beta} (\hat{Q}_{\alpha} \hat{\chi}) \hat{j}^{c}_{\beta} + 
\kappa_{7\alpha ij} (\hat{Q}_{\alpha} \hat{L}_{i}) \hat{d}^{c}_{j} \nonumber \\
&+&
\xi_{1ijk} \hat{d}^{c}_{i} \hat{d}^{c}_{j} \hat{u}^{c}_{k}+
\xi_{2ij\beta} \hat{u}^{c}_{i} \hat{u}^{c}_{j} \hat{j}^{c}_{\beta} +
\xi_{3 i\beta} \hat{d}^{c}_{i} \hat{J}^{c} \hat{j}^{c}_{\beta} . 
\label{sp3m1}
\end{eqnarray}

The SM projection of the 3-3-1 models are into \cite{Pleitez:1993gc}
\begin{equation}
SU(2)_{L}\times U(1)_{Y}\times U(1)_{(B+L)}.	
\label{upa}
\end{equation}
Notice that the $U(1)_{(B+L)}$ factor is a semidirect product.
We can define as in the MSSM the following $R$-Parity 
\cite{Rodriguez:2024rvf,Rodriguez:2010tn}
\begin{equation}
R=(-1)^{2S+3{\cal F}},
\label{rparitymsusy331}
\end{equation} 
where $S$ is the spin and we define the following new quantum number \cite{Rodriguez:2010tn}
\begin{equation}
{\cal F}\equiv B+L
\end{equation}  
where $B$ is the baryon number while $L$ is the total lepton number. The ${\cal F}$ number attribution is \cite{Pleitez:1993gc}
\begin{eqnarray}
{\cal F}(l^{c})&=&-1, \,\ {\cal F}(l) = {\cal F}(\nu_{l})=1, \nonumber \\
{\cal F}(u) &=& - {\cal F}(d)= \frac{1}{3}, \,\ {\cal F}(J)=- \frac{5}{3}, \,\ {\cal F}(j_{1,2})= \frac{7}{3}, \nonumber \\ 
{\cal F}(U^{--})&=&{\cal F}(V^{-}) = {\cal F}(\rho^{++})  = {\cal F}(\eta^{+}_{2})= {\cal F}(H^{++}_{1}) =
{\cal F}(\sigma_{1}^{0})=2. \nonumber \\
{\cal F}(h^{+}_{2})&=& {\cal F}(\chi^{--}) = {\cal F}(H^{--}_{2}) =-2
\label{efe}
\end{eqnarray}  
Choosing the following R-charges 
\begin{eqnarray}
n_{\eta}&=&n_{\rho^{\prime}}=n_{S}=-1, \,\
n_{\rho}=n_{\eta^{\prime}}=n_{S^{\prime}}=1, \,\
n_{\chi}=n_{\chi^{\prime}}=0, \nonumber \\
n_{L}&=&n_{Q_{i}}=n_{d_{i}}=1/2, \,\
n_{J_{i}}=-1/2, \,\ n_{u}=-3/2,
\label{rdiscsusy331} 
\end{eqnarray}
the superpotential of our model that conseves the $R$-Parity, defined in 
Eq.(\ref{rparitymsusy331}), is given by
\begin{eqnarray} 
W^{MSUSY331}_{RPC}&=&\mu_{ \eta} \hat{ \eta} \hat{ \eta}^{\prime}+
\mu_{ \rho} \hat{ \rho} \hat{ \rho}^{\prime}+
\mu_{ \chi} \hat{ \chi} \hat{ \chi}^{\prime}+ 
+
\mu_{S} Tr[(\hat{S} \hat{S}^{\prime})]+
\lambda_{2ij} \epsilon \hat{L}_{iL} \hat{L}_{jL} \hat{ \eta} \nonumber \\
&+&
\lambda_{3ij} 
(\hat{L}_{i}\hat{S}\hat{L}_{j}) +
f_{1} \epsilon \hat{ \rho} \hat{ \chi} \hat{ \eta}+
f_{2} \epsilon \hat{ \chi} \hat{ \rho} \hat{S}+ 
f^{\prime}_{1}\epsilon \hat{ \rho}^{\prime}\hat{ \chi}^{\prime}
\hat{ \eta}^{\prime} +
f^{\prime}_{2} \epsilon \hat{ \chi}^{\prime} \hat{ \rho}^{\prime} 
\hat{S}^{\prime} \nonumber \\ &+&
\kappa_{1\alpha i} \left( \hat{Q}_{\alpha L} \hat{\eta} \right) \hat{d}^{c}_{iL} 
+
\kappa_{2\alpha i} \left( \hat{Q}_{\alpha L} \hat{\rho} \right) \hat{u}^{c}_{iL} +
\kappa_{3\alpha \beta} \left( \hat{Q}_{\alpha L} \hat{\chi} \right) \hat{j}^{c}_{\beta L} 
\nonumber \\ &+&
\kappa_{4i} \left( \hat{Q}_{3 L} \hat{\eta}^{\prime} \right) \hat{u}^{c}_{jL}+
\kappa_{5i} \left( \hat{Q}_{3L} \hat{\rho}^{\prime} \right) \hat{d}^{c}_{iL}+
\kappa_{6 \alpha \beta} \left( \hat{Q}_{\alpha L} \hat{\chi} \right) \hat{j}^{c}_{\beta L} \nonumber \\ &+&
\kappa_{7} \left( \hat{Q}_{3L} \hat{\chi}^{\prime} \right) \hat{J}^{c}_{L}. \nonumber \\
\label{supmsusy331rparitycons}
\end{eqnarray}

In the 331 supersymmetric model the part responsible for the 
supersymmetry breaking can be written as
\begin{equation}
{\cal L}_{\mbox{soft}}={\cal L}_{SMT}+{\cal L}_{GMT}+
{\cal L}^{\mbox{soft}}_{\mbox{scalar}},
\end{equation}
where
\begin{eqnarray} 
{\cal L}_{SMT}&=&{\cal L}^{mass}_{SMT}+{\cal L}^{INT}_{SMT}, \nonumber \\
{\cal L}^{mass}_{SMT}&=&-
 \tilde{L}^{\dagger} m_{\tilde{L}}^{2} \tilde{L}- \sum_{\alpha =1}^{2} 
\tilde{Q}^{\dagger}_{\alpha} m_{\tilde{Q}_{\alpha}}^{2} \tilde{Q}_{\alpha}- 
\tilde{Q}^{\dagger}_{3} m_{\tilde{Q}_{3}}^{2} \tilde{Q}_{3}- \sum_{i=1}^{3} 
\tilde{u}^{c \dagger}_{i}m_{\tilde{u}_{i}}^{2} \tilde{u}^{c}_{i}- \sum_{i=1}^{3}m_{d_{i}}^2 
\tilde{d}^{c \dagger}_{i}m_{\tilde{d}_{i}}^{2} \tilde{d}^{c}_{i} \nonumber \\
&-&  
\tilde{J}^{c \dagger} m_{\tilde{J}}^{2} \tilde{J}^{c} - 
\sum_{\beta =1}^{2}
\tilde{j}^{c \dagger}_{ \beta} m_{\tilde{j}_{ \beta}}^{2} \tilde{j}^{c}_{ \beta}-
m^{2}_{\eta}\eta^{\dagger}\eta -
m^{2}_{\rho}\rho^{\dagger}\rho -
m^{2}_{\chi}\chi^{\dagger}\chi -
m^{2}_{S}S^{\dagger}S \nonumber \\
&-&
m^{2}_{\eta^{\prime}}\eta^{\prime \dagger}\eta^{\prime}-
m^{2}_{\rho}\rho^{\prime \dagger}\rho^{\prime}-
m^{2}_{\chi}\chi^{\prime \dagger}\chi^{\prime}
-m^{2}_{S}S^{\prime \dagger}S^{\prime}, 
\end{eqnarray}
This term gives mass to the sleptons, squarks\footnote{Tthe partners of 
leptons and quarks respectivelly} and usual scalars, where
\begin{equation}
m_{\tilde{L}}^{2} = \left( \begin{array}{ccc} 
m^{2}_{\tilde{\nu}}& 0& 0 \\ 
0& m^{2}_{\tilde{l}}& 0 \\
0& 0&  m^{2}_{\tilde{l}^{c}}
\end{array} \right),  
m_{\tilde{Q}_{\alpha}}^{2} = \left( \begin{array}{ccc} 
m^{2}_{\tilde{d}_{\alpha}}& 0& 0 \\ 
0& m^{2}_{\tilde{u}_{\alpha}}& 0 \\
0& 0&  m^{2}_{\tilde{j}_{\alpha}}
\end{array} \right), 
m_{\tilde{Q}_{3}}^{2} = \left( \begin{array}{ccc} 
m^{2}_{\tilde{u}_{3}}& 0& 0 \\ 
0& m^{2}_{\tilde{d}_{3}}& 0 \\
0& 0&  m^{2}_{\tilde{J}}
\end{array} \right).
\label{mm}
\end{equation}

The interactions between the scalars are described by the following term
\begin{eqnarray} 
{\cal L}^{INT}_{SMT}&=&
-A_{1i} \tilde{L}_{i} \eta^{\prime}+ A_{21j} \tilde{L}_{i} \tilde{L}_{j} \eta  
A_{31j}  \tilde{L}_{i} \tilde{L}_{j} S_{ij}+ 
\tilde{Q}^{\dagger}_{1}( \zeta_{1i}  \eta \tilde{u}^{c}_{i}+
\zeta_{2i}  \rho \tilde{d}^{c}_{i}+ \zeta_{3J} \chi \tilde{J}^{c})  
\nonumber \\
&+&   
\tilde{Q}^{\dagger}_{\alpha} \left[ (\omega_{1\alpha i} \eta \tilde{d}^{c}_{i} + 
\omega_{2\alpha i} \rho \tilde{u}^{c}_{i}+
\omega_{3 \alpha ij}  \tilde{L}_{i} \tilde{d}^{c}_{j}) +
\omega_{4\alpha \beta}  \chi \tilde{j}^{c}_{\beta} \right] \nonumber \\  
&+&  \varsigma_{1ijk} \tilde{d}^{c}_{i} \tilde{d}^{c}_{j} \tilde{u}^{c}_{k}+ 
\varsigma_{2i\beta} \tilde{d}^{c}_{i} \tilde{J}^{c} \tilde{j}^{c}_{\beta}+
\varsigma_{3ij\beta} \tilde{u}^{c}_{i} \tilde{u}^{c}_{j} \tilde{j}^{c}_{\beta} +
h_1\epsilon_{ijk}\rho_i\chi_j\eta_k+h_2\eta_i\eta_j \bar{S}_{ij} \nonumber \\
&+&h_3\chi_i\rho_j \bar{S}_{ij}+hc . 
\label{smt}
\end{eqnarray}
This term gives the interactions with the scalars.

If we consider the term $\hat{L}_{i}\hat{\eta}^{\prime}$, in our superpotential defined in 
Eq.(\ref{supmsusy331rparitycons})\footnote{See also Eq.(\ref{w2geral}).}, 
it will generate
\begin{eqnarray}
\mu_{0i}(\hat{L}_{i}\hat{\eta}^{\prime}) = \mu_{0i} \left( \nu_{i}\tilde{\eta}^{\prime 0}+
l_{i}\tilde{\eta}^{+}_{1}+ l^{c}_{i}\tilde{\eta}^{-}_{2} \right),
\end{eqnarray}
it induce the mixing between usual leptons with the usual higgsinos as we discussed in \cite{Montero:2001ch}. 
Recently, we consider the masses of fermions in the context of $R$-Parity conservation\cite{Rodriguez:2025vea}, next 
we want to consider scenarios with $R$-Parity violation. We have also to the following term
\begin{eqnarray}
A_{1i} \tilde{L}_{i} \eta^{\prime}=A_{1i} \left( \tilde{\nu}_{i}\eta^{\prime 0}+
\tilde{l}_{i}\eta^{+}_{1}+ \tilde{l}^{c}_{i}\eta^{-}_{2} \right),
\end{eqnarray} 
this term induce mixing with sleptons and usual scalars, we should include it in a future work because it was not considered 
in our previous analyses presented in \cite{Rodriguez:2024rvf}.

 The second term is
\begin{equation}
{\cal L}_{GMT}={\cal L}^{\mbox{gluino}}_{GMT}+{\cal L}^{\mbox{gaugino}}_{GMT},
\end{equation}
the first piece is
\begin{equation}
{\cal L}^{\mbox{gluino}}_{GMT}=- \frac{1}{2} m_{ \lambda_{C}} \sum_{a=1}^{8} 
\left( \lambda^{a}_{C} \lambda^{a}_{C}+hc \right),
\end{equation}
this is the mass term to the gluinos, partners of the gluons. The second 
part is
\begin{equation}
{\cal L}^{\mbox{gaugino}}_{GMT}=- \frac{1}{2} \left[ 
m_{ \lambda} \sum_{a=1}^{8} 
\left( \lambda^{a}_{A} \lambda^{a}_{A} \right) 
+m^{ \prime} \lambda_{B} \lambda_{B}+hc \right].
\label{gmt}
\end{equation}
which gives mass to the gauginos, partners of the gauge bosons.

\section{Sleptons}

The diagonal mass term for the sleptons are
\begin{eqnarray} 
- \tilde{L}^{\dagger}_{i} \left( m_{\tilde{L}}^{2} \right)_{ij} \tilde{L}_{j}=- \left[
\left( m^{2}_{\tilde{ \nu}} \right)_{ij} \tilde{ \nu}^{*}_{i} \tilde{ \nu}_{j}+
\left( m^{2}_{l} \right)_{ij} \tilde{l}^{*}_{i} \tilde{l}_{j}+
\left( m^{2}_{\tilde{l}^{c}} \right)_{ij} \tilde{l}^{c*}_{i} \tilde{l}^{c}_{j} \right],
\end{eqnarray}
such as in the MSSM in this term above there are terms allow flavor changing neutral currents. To avoid it we will consider 
diagonal form for $ m_{\tilde{L}}^{2}$, it means $\left( m_{\tilde{L}}^{2} \right)_{ii}$. The mixing between the sleptons are given by
\begin{eqnarray}
{\cal L}^{INY}&=&A_{2ij} \left( \tilde{L}_{i}\tilde{L}_{j}\eta \right) + A_{3ij}
\left( \tilde{L}_{i} \tilde{L}_{j} S \right) +hc \nonumber \\ &=&
\frac{A_{2ij}v_{\eta}}{\sqrt{2}}\left( \tilde{l}_{i}\tilde{l}^{c}_{j}- \tilde{l}^{c}_{i}\tilde{l}_{j} \right) +
\frac{A_{3ij}}{\sqrt{2}}\left[ z_{1} \tilde{\nu}_{i}\tilde{\nu}_{j}+ 
z_{2}\left( \tilde{l}_{i}\tilde{l}^{c}_{j}- \tilde{l}^{c}_{i}\tilde{l}_{j} \right) \right].
\end{eqnarray}

The another two parts that contribute to the sleptons mass are the F-terms and 
the D-terms, and we get the mass contributions from these equations
\begin{eqnarray}
{\cal L}^{\mbox{slepton}}_{F}&=&- \left[ \left| 
2 \lambda_{2} \left(  \epsilon_{ijk} \tilde{L}_{j} \eta_{k} \right) +
2 \lambda_{3} S_{ij}\tilde{L}_j \right|^2 \right. \nonumber \\
&+& \left. \left| \mu_{ \eta} \eta^{\prime}_i+
\lambda_{2} \left( \epsilon_{ijk} \tilde{L}_{j} \tilde{L}_{k} \right) +
f_{1} \left( \epsilon_{ijk} \rho_{j} \chi_{k} \right) +
2f_{2} S_{ij}\eta_j \right|^2 \right. \nonumber \\
&+& \left. \left|\mu_{S} S^{\prime}_{ij}+
f_{2} \eta_i \eta_j+
\lambda_{3} \tilde{L}_i \tilde{L}_j+
f_{3} \rho_i \chi_j \right|^2 \right], \nonumber\\
\epsilon_{kij}\epsilon_{klm}&=& \delta_{il}\delta_{jm}- \delta_{im}\delta_{jl}, \nonumber \\
{\cal L}^{\mbox{slepton}}_D&=&- \frac{g^2}{8} \left(  
\eta^{*}_i \lambda^{a}_{ij} \eta_j+ \rho^{*}_i \lambda^{a}_{ij} \rho_j+
\chi^{*}_i \lambda^{a}_{ij} \chi_j-S^{*}_{ij} \lambda^{*a}_{jk}S_{kl}-
\eta^{\prime *}_i \lambda^{*a}_{ij} \eta^{\prime}_j- \rho^{\prime *}_i \lambda^{*a}_{ij} \rho^{\prime}_j 
\right. \nonumber \\ &-& \left.
\chi^{\prime *}_i \lambda^{*a}_{ij} \chi^{\prime}_j+S^{\prime *}_{ij} \lambda^{a}_{jk}S^{\prime}_{kl}+
\tilde{L}^{*}_i  \lambda^a_{ij} \tilde{L}_j \right)^{2} \nonumber \\
\lambda^{a}_{ij}\lambda^{a}_{kl}&=&2 \delta_{il}\delta_{jk}- \frac{2}{3} \delta_{ij}\delta_{kl}.
\end{eqnarray}
developing these terms and using Eqs.(\ref{sl},\ref{vev},\ref{vev1}) we get
\begin{eqnarray}
{\cal L}^{\mbox{slepton}}_F&=&- \frac{4}{9} \left[ \lambda^2_2v^2
(\tilde{l}^* \tilde{l}+ \tilde{l}^{c *} \tilde{l}^c)+ \lambda^2_3 \left( z^2_1 
\tilde{ \nu}^*_l \tilde{ \nu}_l+ \frac{z^2}{2}( \tilde{l}^* \tilde{l}+ 
\tilde{l}^{c *} \tilde{l}^c) \right. \right. \nonumber \\
&+& \left. \left. \frac{9 \lambda_{3}}{4} \left( 
\frac{\mu_Sz}{\sqrt{2}}+ \frac{f_3uw}{9} \right) (\tilde{l} \tilde{l}^{c}+
\tilde{l}^{*} \tilde{l}^{c *}) \right) \right], \nonumber \\
{\cal L}^{\mbox{slepton}}_D&=&- \frac{g^2}{2} \left[ 
\left( \frac{z^2+4v^2+4u^2+4w^2+4z^2_1}{3} \right) \tilde{ \nu}^*_l 
\tilde{ \nu}_l+
\left( \frac{5z^2+2z^2_1}{3} \right) ( \tilde{l}^* \tilde{l}+ 
\tilde{l}^{c *} \tilde{l}^c)
\right].
\label{smt2}
\end{eqnarray}

\subsection{Charged Sleptons}

We can write the slepton mass term as
\begin{eqnarray} 
\frac{2 \lambda^2_3z^2}{9} \tilde{l}^* \tilde{l}- \left( \tilde{m}^2_R+ 
\frac{2 \lambda^2_3z^2}{9}  \right) \tilde{l}^{c *} \tilde{l}^c+ 
\tilde{m}^2_{LR}( \tilde{l} \tilde{l}^{c}- \tilde{l}^{c *} \tilde{l}^*),
\label{smt3}  
\end{eqnarray}
where
\begin{eqnarray}
\tilde{m}^2_L&=&m^2_l+\lambda^{2}_{2}v^{2}+ 
\frac{z^2-2(z^2_1+w^2-v^2-u^2)}{12}, \nonumber \\
\tilde{m}^2_R&=&m^2_{l^c}+\lambda^{2}_{2}v^{2}+ 
\frac{z^2-2(z^2_1-2w^2-v^2+u^2)}{12}, 
\nonumber \\
\tilde{m}^2_{LR}&=& \frac{9 \lambda_{3}}{4} \left( \frac{ \mu_Sz}{\sqrt{2}}+ 
\frac{f_3uw}{9} \right)- \zeta_0z.
\end{eqnarray}

Performing the diagonalization we find the following states
\begin{eqnarray}
&&\left( \begin{array}{c}
\tilde{l}^-_1 \\
\tilde{l}^-_2
\end{array} \right)= \left( \begin{array}{cc}
\cos \theta_f& \sin \theta_f \\
- \sin \theta_f& \cos \theta_f
\end{array} \right) \left( \begin{array}{c}
\tilde{l} \\
\tilde{l}^{c *}
\end{array}\right), \nonumber \\
&&\left( \begin{array}{cc}
\tilde{l}^+_1& \tilde{l}^+_2
\end{array} \right)=
\left( \begin{array}{cc}
\tilde{l}^{*}& \tilde{l}^{c}
\end{array}\right) 
\left( \begin{array}{cc}
\cos \theta_f&- \sin \theta_f \\
\sin \theta_f& \cos \theta_f
\end{array} \right),
\label{defslepmass}
\end{eqnarray}
where the mixing angle is given by
\begin{equation}
\tan 2 \theta_f= \frac{2 \tilde{m}^2_{LR}}{ \tilde{m}^2_L- \tilde{m}^2_R},
\label{mixingangleslep}
\end{equation}
and the mass of these states are
\begin{equation}
m^2_{1,2}= \frac{2 \lambda^2_3z^2}{9}+ \frac{1}{2} \left[ \tilde{m}^2_L+ 
\tilde{m}^2_R \pm \sqrt{ ( \tilde{m}^2_L- \tilde{m}^2_R)^2+4 \tilde{m}^4_{LR}}
\right].
\end{equation}

Substituing the diagonal sleptons states, given in Eq.(\ref{defslepmass}), in 
the Eq.(\ref{smt3}), we obtain the following diagonal lagrangian
\begin{equation}
{\cal L}^{ \mbox{slepton}}_{ \mbox{mass}}=- \tilde{m}_{\tilde{ \nu}} 
\tilde{ \nu}^*_l \tilde{ \nu}_l- m^2_1 \tilde{l}^+_1 \tilde{l}^-_1-
m^2_2 \tilde{l}^+_2 \tilde{l}^-_2.
\end{equation}
We wish to stress that the slepton sector of this model is the same as in the 
MSSM \cite{mssm}.

\subsection{Neutral Sleptons}

\begin{eqnarray}
- \tilde{m}^{2}_{\tilde{ \nu}} \tilde{ \nu}^*_{l} \tilde{ \nu}_{l}.
\label{smt3}  
\end{eqnarray}

\section{Conclusions}
\label{sec:conclusion}
We have studied the fermions masses of supersymmetric 
331 model with the sextet and anti-sextet and our results are in 
agreement with the actual experimental data. Here we consider 
only the anti-sextet $S$ get VEV. We want to extend this analysis by 
also allowing the sextet $S^{\prime}$ to acquire vev, in similar way as 
we have done recently \cite{Rodriguez:2022hsj,Rodriguez:2024rvf}. 

\begin{center}
{\bf Acknowledgments} 
\end{center}
We would like thanks V. Pleitez for useful discussions above 331 models. We 
also to thanks IFT for the nice hospitality during my several visit 
to perform my studies about the severals 331 Models and also for done 
this article. 

\section{Minimal Supersymmetric $331$ Model.}
\label{msusy331}



\begin{thebibliography}{99}
\bibitem{singer}M. Singer, J. W. F. Valle, and J. Schechter, 
{\it Canonical neutral-current predictions from the weak-electromagnetic gauge group $SU(3)\otimes U(1)$}, 
{\sl Phys. Rev.} {\bf D 22}, 738, (1980).
\bibitem{ppf}F. Pisano and V. Pleitez, 
{\it An $SU(3)\otimes U(1)$ model for electroweak interactions}, 
{\sl Phys. Rev.}  {\bf D 46}, 410, (1992); 
R. Foot, O. F. Hernandez, F. Pisano and V. Pleitez, 
{\it Lepton masses in an $SU(3)_{L} \otimes U(1)_{N}$ gauge model},
{\sl Phys. Rev.} {\bf D 47}, 4158, (1993).
\bibitem{331rh} R. Foot, H.N. Long and Tuan A. Tran,
{\it $SU(3)_{L}\otimes U(1)_{N}$ and $SU(4)_{L}\otimes U(1)_{N}$ gauge models with right-handed neutrinos}, 
{\sl Phys. Rev.} {\bf D 50}, 34, (1994);
J. C. Montero, F. Pisano and V. Pleitez, {\sl Phys. Rev.} {\bf D 47}, 2918, (1993); 
H.N. Long, 
{\it $SU(3)_{L}\otimes U(1)_{N}$ model for right-handed neutrino neutral currents}, 
{\sl Phys. Rev.} {\bf D 54}, 4691, (1996); 
H.N. Long,
{\it $SU(3)_{C}\otimes SU(3)_{L}\otimes U(1)_{N}$ model with right-handed neutrinos}, 
{\sl Phys. Rev.} {\bf D 53}, 437, (1996).
\bibitem{Pleitez:1992xh} V. Pleitez and M. D. Tonasse,
{\it Heavy charged leptons in an SU(3)-L x U(1)-N model},
{\sl Phys. Rev.}{\bf D48}, 2353, (1993); [arXiv:hep-ph/9301232 [hep-ph]].
\bibitem{Pleitez:1993gc} V. Pleitez and M. D. Tonasse,
{\it Neutrinoless double beta decay in an SU(3)-L x U(1)-N model},''
{\sl Phys. Rev.}{\bf D48}, 5274, (1993);  [arXiv:hep-ph/9302201 [hep-ph]].
\bibitem{Pleitez:1994pu} V. Pleitez,
{\it New fermions and a vector - like third generation in SU(3) (C) x SU(3) (L) x U(1) ($N$) models},
{\sl Phys. Rev.}{\bf D53}, 514, (1996); [arXiv:hep-ph/9412304 [hep-ph]].
\bibitem{Ponce:2001jn} W. A. Ponce, J. B. Florez and L. A. Sanchez,
{\it Analysis of $SU(3)(c) x SU(3)(L) x U(1)(X)$ local gauge theory},
{\sl Int. J. Mod. Phys.}{\bf A17}, 643, (2002); [arXiv:hep-ph/0103100 [hep-ph]].
\bibitem{Diaz:2004fs} R. A. Diaz, R. Martinez and F. Ochoa,
{\it SU(3)(c) x SU(3)(L) x U(1)(X) models for beta arbitrary and families with mirror fermions},
{\sl Phys. Rev.}{\bf D72}, 035018, (2005); [arXiv:hep-ph/0411263 [hep-ph]].
\bibitem{Pleitez:2021abk} V. Pleitez,
{\it Challenges for the 3-3-1 models}, [arXiv:2112.10888 [hep-ph]].
\bibitem{Long:2024gyy} H. N. Long,
{\sl New Physics in the 3-3-1 models},
{\it Commun. in Phys.}{\bf 34}, 203, (2024); [arXiv:2412.19188 [hep-ph]].
\bibitem{Sanchez:2001ua}L. A. Sanchez, W. A. Ponce and R. Martinez,
{\it SU(3) ($c$) x SU(3) ($\ell$) x U(1) ($X$) as an E(6) subgroup},
{\sl Phys. Rev.} {\bf D64}, 075013, (2001).
\bibitem{Martinez:2001mu}
R. Martinez, W. A. Ponce and L. A. Sanchez,
{\it SU(3) (C) x SU(3) (L) x U(1) ($X$) as an SU(6) x U(1) ($X$) subgroup},
{\sl Phys. Rev.} {\bf D65}, 055013, (2002).
\bibitem{Martinez:2011iq} R. Martinez, F. Ochoa and P. Fonseca,
{\it A 3-3-1 Model with SU(8) Unification}, [arXiv:1105.4623 [hep-ph]].
\bibitem{Frampton:1989fu} P. H. Frampton and B. H. Lee,
{\it SU(15) GRAND UNIFICATION},
{\sl Phys. Rev. Lett.}{\bf 64}, 619, (1990).
\bibitem{Cuypers:1996ia}F. Cuypers and S. Davidson,
{\it Bileptons: Present limits and future prospects},
{\sl Eur. Phys. J.}{\bf C2}, 503, (1998).
\bibitem{Liu:1993gy}
J. T. Liu and D. Ng,
\emph{Lepton flavor changing processes and CP violation in the 331 model},
\emph{ Phys. Rev.}{\bf D50}, 548, (1994).
\bibitem{Liu:1993fwa} J. T. Liu and D. Ng,
\emph{Z - Z-prime mixing and oblique corrections in an SU(3) x U(1) model},
\emph{Z. Phys.}{\bf C62}, 693, (1994).
\bibitem{Montero:1999mc}
J. C. Montero, C.A. de Sousa Pires and V. Pleitez,
{\it Spontaneous breaking of a global symmetry in a 331 model},
{\sl Phys. Rev.}{\bf D60}, 115003, (1999).
\bibitem{DeConto:2015eia} G. De Conto, A. C. B. Machado and V. Pleitez,
{\it Minimal 3-3-1 model with a spectator sextet},
{\sl Phys. Rev.}{\bf D92}, 075031, (2015).
\bibitem{Machado:2010uc} A. C. B. Machado, J. C. Montero and V. Pleitez,
{\it Three-Higgs-doublet model with $A_4$ symmetry},
{\sl Phys. Lett.}{\bf B697} 318 (2011); [arXiv:1011.5855 [hep-ph]].
\bibitem{Zee:1980ai}A. Zee,
{\it A Theory of Lepton Number Violation, Neutrino Majorana Mass, and Oscillation},
{\sl Phys. Lett.}{\bf B93}, 389, (1980); [erratum: 
{\sl Phys. Lett.}{\bf B95}, 461, (1980)].
\bibitem{Machado:2017flo}
A. C. B. Machado, J. Monta\~no, P. Pasquini and V. Pleitez,
{\it Analytical solution for the Zee mechanism}, [arXiv:1707.06977 [hep-ph]].
\bibitem{Zee:1985id} A. Zee,
{\it Quantum Numbers of Majorana Neutrino Masses},
{\sl Nucl. Phys.}{\bf B264}, 99, (1986).
\bibitem{Babu:1988ki} K. S. Babu,
{\it Model of 'Calculable' Majorana Neutrino Masses},
{\sl Phys. Lett.}{\bf B203}, 132, (1988).
\bibitem{Gelmini:1980re} G. B. Gelmini and M. Roncadelli,
{\it Left-Handed Neutrino Mass Scale and Spontaneously Broken Lepton Number},
{\sl Phys. Lett.}{\bf B99}, 411 (1981).
\bibitem{Schechter:1980gr} J. Schechter and J. W. F. Valle,
{\it Neutrino Masses in SU(2) x U(1) Theories},
{\sl Phys. Rev.}{\bf D22}, 2227, (1980).
\bibitem{Cheng:1980qt} T. P. Cheng and L. F. Li,
{\it Neutrino Masses, Mixings and Oscillations in SU(2) x U(1) Models of Electroweak Interactions},
{\sl Phys. Rev.}{\bf D22}, 2860, (1980).
\bibitem{Aoki:2011yk} M. Aoki, S. Kanemura and K. Yagyu,
{\it Doubly-charged scalar bosons from the doublet},
{\sl Phys. Lett.}{\bf B702}, 355, (2011);
[erratum: {\sl Phys. Lett.}{\bf B706}, 495, (2012)];
\bibitem{Puerta:2021hfl} M. Puerta,
{\it Dark matter in a charged variant of the Scotogenic model},
[arXiv:2110.09805 [hep-ph]].

\bibitem{deSousaPires:1998jc} C. A. de Sousa Pires and O. P. Ravinez,
{\it Charge quantization in a chiral bilepton gauge model},
{\sl Phys. Rev.}{\bf D58}, 035008, (1998); [arXiv:hep-ph/9803409 [hep-ph]].
\bibitem{Dong:2005ebq} P. V. Dong and H. N. Long,
{\it Electric charge quantization in SU(3)(C) x SU(3)(L) x U(1)(X) models},
{\sl Int. J. Mod. Phys.}{\bf A21}, 6677, (2006); [arXiv:hep-ph/0507155 [hep-ph]].
\bibitem{pal1} P. B. Pal,  
{\it The Strong CP question in SU(3)(C) x SU(3)(L) x U(1)(N) models},
\emph{Phys. Rev.}{\bf D52}, 1659, (1995). 
\bibitem{Montero:1998nf}J. C. Montero, V. Pleitez and O. Ravinez,
\emph{Concerning CP violation in 331 models},
\emph{AIP Conf. Proc.}{\bf 445}, 353, (1998).
\bibitem{Montero:1998yw}J. C. Montero, V. Pleitez and O. Ravinez,
\emph{Soft superweak CP violation in a 331 model},
\emph{Phys. Rev.}{\bf D60}, 076003, (1999).
\bibitem{Montero:2005yb}
J. C. Montero, C. C. Nishi, V. Pleitez, O. Ravinez and M. C. Rodriguez,
\emph{Soft CP violation in $K$ meson systems},
\emph{Phys. Rev.}{\bf D73}, 016003, (2006).
\bibitem{ema1} T. V. Duong and E. Ma, 
{\it Supersymmetric $SU(3) \otimes U(1)$ Gauge Model: 
Higgs Structure at the Electroweak Energy Scale}, 
{\sl Phys. Lett.}{\bf B316}, 307 (1993).
\bibitem{pal2} H. N. Long and P. B. Pal, {\it Nucleon instability in a supersymmetric 
$SU(3)_{C}\otimes SU(3)_{L}\otimes U(1)$ model}, 
{\sl Mod. Phys. Lett.}{\bf A13}, 2355, (1998).
\bibitem{331susy1}J. C. Montero, V. Pleitez and M. C. Rodriguez, 
{\it A Supersymmetric 3-3-1 model},
{\sl Phys. Rev.} {\bf D65}, 035006, (2002).
\bibitem{mcr} M. Capdequi-Peyran\`ere and M.C. Rodriguez, 
{\it Charginos and neutralinos production at 3-3-1 supersymmetric model in e- e- scattering},
{\sl Phys. Rev.} {\bf D 65}, 035001 (2002);
M. C. Rodriguez,
{\it Double Chargino Production in $e^{-}e^{-}$ scattering},
{\sl Int. J. Mod. Phys. }{\bf A 22}, 6080, (2008).
\bibitem{Rodriguez:2005jt}M. C. Rodriguez,
{\it Scalar sector in the minimal supersymmetric 3-3-1 model},
{\sl Int. J. Mod. Phys.}{\bf A21}, 4303, (2006).
\bibitem{Rodriguez:2022hsj}M. C. Rodriguez,
{\it Gauge bosons masses in the context of the Supersymmetric $SU(3)_{C}\otimes SU(3)_{L}\otimes U(1)_{N}$ Model},
{\sl Int. J. Mod. Phys.}{\bf A}, 2440001, (2024);
[arXiv:2209.04653 [hep-ph]].
\bibitem{Rodriguez:2024rvf} M. C. Rodriguez,
{\it Spontaneous breaking of a global symmetry in the Minimum Supersymmetric $SU(3)_{C}\times SU(3)_{L}\times U(1)_{N}$ Model},
[arXiv:2408.08695 [hep-ph]].
\bibitem{Rodriguez:2025vea} M. C. Rodriguez,
{\it The Masses of Fermions in the context of the Supersymmetric $SU(3)_{C}\times SU(3)_{L}\times U(1)_{N}$ Model},
[arXiv:2508.11456 [hep-ph]].
\bibitem{Ng:1992st} D. Ng,
{\it The Electroweak theory of $SU(3) x U(1)$},
{\sl Phys. Rev.}{\bf D49}, 4805, (1994); [arXiv:hep-ph/9212284 [hep-ph]].
\bibitem{Rodriguez:2010tn} M. C. Rodriguez,
{\it Mass Spectrum in the Minimal Supersymmetric 3-3-1 model},
{\sl J. Mod. Phys.}{\bf 2}, 1193, (2011); [arXiv:1007.0981 [hep-ph]].
\bibitem{DeConto:2017aob} G. De Conto and V. Pleitez,
{\it Hadronic resonances with exotic electric charge},
{\sl Phys. Rev.}{\bf D96}, 075028, (2017); [arXiv:1708.06285 [hep-ph]].
\bibitem{dumm2} D. G. Dumm,
{\it Leptophobic character of the Z-prime in an SU(3)(C) x SU(3)(L) x U(1)(X) model},
{\sl Phys. Lett.}{\bf B411}, 313, (1997); [arXiv:hep-ph/9709245 [hep-ph]].
\bibitem{Buras:2012dp} A. J. Buras, F. De Fazio, J. Girrbach and M. V. Carlucci,
{\it The Anatomy of Quark Flavour Observables in 331 Models in the Flavour Precision Era},
{\it JHEP}{\bf 02}, 023 (2013); [arXiv:1211.1237 [hep-ph]].
\bibitem{Buras:2014yna} A. J. Buras,  F. De Fazio and J. Girrbach-Noe,
{\it $Z$-$Z'$ mixing and $Z$-mediated FCNCs in $SU(3)_{C}  \times  SU(3)_{L}  \times U(1)_{X}$ models},
{\sl JHEP}{\bf 08}, 039, (2014); [arXiv:1405.3850 [hep-ph]].
\bibitem{Buras:2021rdg} A. J. Buras, P. Colangelo, F. De Fazio and F. Loparco,
{\it The charm of 331},
{\sl JHEP}{\bf 10}, 021, (2021); [arXiv:2107.10866 [hep-ph]].
\bibitem{Montero:2001ch} J. C. Montero, V. Pleitez and M. C. Rodriguez,
{\it Lepton masses in a supersymmetric 3-3-1 model},
{\sl Phys. Rev}{\bf. D65}, 095008, (2002); [arXiv:hep-ph/0112248 [hep-ph]].



\end{thebibliography}
\end{document}